\documentclass[twocolumn,tighten]{aastex631}

\usepackage{amsmath}

\begin{document}

\title{The Impact of an AGN on PAH Emission in Galaxies: the Case of Ring Galaxy NGC\,4138}

\author[0009-0001-6065-0414]{G. P. Donnelly}
\affiliation{Ritter Astrophysical Research Center, University of Toledo, Toledo, OH 43606, USA}

\author[0000-0003-1545-5078]{J.D.T. Smith}
\affiliation{Ritter Astrophysical Research Center, University of Toledo, Toledo, OH 43606, USA}

\author[0000-0002-0846-936X]{B. T. Draine}
\affiliation{Dept. of Astrophysical Sciences, Princeton University, Princeton, NJ 08544, USA}

\author[0000-0001-5042-3421]{A. Togi}
\affiliation{Dept. of Physics, Texas State University, 601 University Drive, San Marcos, TX 78666, USA}

\author[0000-0001-8490-6632]{T. S.-Y. Lai}
\affiliation{IPAC, California Institute of Technology, 1200 E. California Blvd., Pasadena, CA 91125, USA}

\author[0000-0003-3498-2973]{L. Armus}
\affiliation{IPAC, California Institute of Technology, 1200 E. California Blvd., Pasadena, CA 91125, USA}

\author[0000-0002-5782-9093]{D. A. Dale}
\affiliation{Dept. of Physics \& Astronomy, University of Wyoming, Laramie, WY, 82071, USA}

\author[0000-0002-2688-1956]{V. Charmandaris}
\affiliation{Dept. of Physics, University of Crete, Heraklion, 71003, Greece}
\affiliation{Institute of Astrophysics, Foundation for Research and Technology-Hellas (FORTH), Heraklion, 70013, Greece}
\affiliation{School of Sciences, European University Cyprus, Diogenes street, Engomi, 1516 Nicosia, Cyprus}

\begin{abstract}

We present a focused study of radially-resolved varying PAH emission in the low-luminosity AGN-host NGC\,4138 using deep \textit{Spitzer}/IRS spectral maps. Using new model PAH spectra, we investigate whether these variations could be associated with changes to the PAH grain size distribution due to photodestruction by the AGN. Separately, we model the effects of the varying radiation field within NGC\,4138, and we use this model to predict the corresponding changes in the PAH emission spectrum. We find that PAH band ratios are strongly variable with radius in this galaxy with short-to-long wavelength band ratios peaking in the starburst ring. The changing mix of starlight appears to have a considerable effect on the trends in these band ratios, and our radiation model predicts the shapes of these trends. However, the amplitude of observed variation is $\mathrm{\sim2.5\times}$ larger than predicted for some ratios. A cutoff of small grains in the PAH size distribution, as has been suggested for AGN, together with changes in PAH ionization fraction could explain the behavior of the shorter bands, but this model fails to reproduce longer band behaviors. Additionally, we find that short-to-long wavelength PAH band ratios increase slightly within $\mathrm{\sim270\,pc}$ of the center, suggesting that the AGN may directly influence PAH emission there.

\end{abstract}


\section{Introduction} \label{sec:intro}

The presence of an active galactic nucleus (AGN) has been shown to have significant impacts on the properties of the interstellar medium (ISM) of a galaxy. Recently, particular attention has been paid to small dust grains in the ISM known as polycyclic aromatic hydrocarbons (PAHs), which glow as prominent emission features in the mid-infrared (MIR) of most galaxies. MIR surveys of galaxies have provided circumstantial evidence that AGN may influence the emission of these PAHs, although the mechanism responsible for this is uncertain. 

It has been known for decades that PAH features are significantly suppressed or even non-existent in the spectra of galaxies with a sufficiently powerful AGN \citep{roche_91, xie_22}. \cite{voit_92} used observations of suppressed PAH features along with laboratory experiments to argue that the hard X-ray emission from an AGN will destroy PAH grains nearby. 

In spite of this, it has become clear that PAHs are present in galaxies with low to moderate luminosity AGN, even within the central 10's of pc \citep{alonso-herrero_14, jensen_17}. \cite{smith_07} (hereafter S07) detected suppressed overall PAH emission in low-luminosity AGN (LLAGN), but also pointed out that the relative strengths of different PAH features are different in the inner $\sim$kpc of LLAGN hosts compared to the galaxies with no AGN (hereafter ``normal"), a result that has been confirmed several times \citep{odowd_09, diamond-stanic_10, sales_10}. PAH emission models suggest that  the relative strengths of PAH features are heavily influenced by the size distribution of PAH grains \citep{draine_07, draine_21}, leading S07 to suggest that an LLAGN may be able to influence this grain size distribution (GSD) by selectively destroying smaller grains. A confirmed influence of an LLAGN on the PAH GSD would be a major result, having implications for the way that PAH features are used as diagnostics of gas, dust, and star-formation within galaxies and potentially allowing for the use of PAH observations as a diagnostic for an AGN itself.

Other factors are known to influence the relative strengths of PAH features, and these may account for the variability observed in PAH band ratios in the absence of an AGN. For example, the ionization state of PAHs is usually studied alongside PAH size because of the characteristic tendency of more ionized PAHs to emit more strongly in certain features. For the \textit{ISO} and \textit{Spitzer} spectra of a large sample of Milky Way and extragalactic systems of different types, \cite{galliano_08} attributed the observed band ratio variability as primarily due to this. Similarly, band ratios are known to vary with the gas-phase metallicity of the ISM, though the exact reason for this is unknown (\cite{sandstrom_2012}, Whitcomb et al. in prep.). The intensity and average photon energy of the interstellar radiation field (ISRF) incident on PAHs is also expected to have impacts on the relative strengths of PAH features by changing the temperature of PAH grains \citep{draine_21}.

In particular, the varying ISRF has been mostly overlooked in investigations of PAH band ratios in LLAGN host galaxies. The ISRF is not uniform across a galaxy, and is linked to galaxy properties such as the presence of an AGN and the age and distribution of the predominant local stellar populations \citep{dale23}. Though it is usually assumed that the UV-rich ISRFs of young stellar populations are responsible for the majority of PAH excitation, observational evidence confirms that UV-poor older stellar populations are also effective for PAH excitation \citep{groves_12, crocker_13, bendo_20, zhang_23}. This leads to the possibility that LLAGN hosts have distinct PAH band ratios not because of direct influence on PAHs from the LLAGN, but because the classification of an AGN is associated with a lower star-formation rate and a larger bulge within a galaxy.

Can the PAH band ratios of a LLAGN host be reproduced by varying the radiation incident upon PAHs alone? If not, does LLAGN-driven small grain destruction better explain the band ratios? New models from \cite{draine_21} (hereafter D21) predict PAH emission spectra given different incident spectra across a range of intensity and hardness. In this work, we use the D21 models to predict the radial band ratio profiles for the LLAGN host NGC\,4138 for a fixed PAH GSD and ionization distribution, and we compare these to the band ratios observed in a deep \textit{Spitzer} spectral map. We then discuss whether these observed band ratios could be explained by alterations to the GSD by the LLAGN.

\section{PAH Band Ratios in AGN Galaxies} \label{background}

PAH features are complexes of emission from the discrete vibrational modes of excited PAH molecules. Thus, the relative strengths of PAH features depend on intrinsic properties of the grain population. Models suggest that the 6.2\,$\mathrm{\micron}$ and 7.7\,$\mathrm{\micron}$ features are primarily emitted by smaller PAHs (see \cite{draine_07} and \cite{draine_21}). This is the result of smaller PAH grains having smaller heat capacities; given the same incident radiation, smaller PAHs will reach the higher temperatures needed to radiate at shorter wavelengths. On the other hand, the 11.3\,$\micron$ and 17\,$\micron$ features are thought to arise primarily from larger PAH grains, given the larger heat capacities of the grains. The 3.3\,$\micron$ feature is thought to be the most sensitive to PAH size \citep{lai_20, draine_21, maragkoudakis_2022, sandstrom_23}, however this feature unfortunately lies outside of the range of \textit{Spitzer}/IRS. 

Given the dependence of different features on different parameters, it has become common for ratios of PAH feature strengths to be used as diagnostics for the grain population. Many authors have used PAH feature ratio-ratio diagrams to probe PAH size, typically using $\mathrm{L(6.2\,\mu m)/L(7.7\,\mu m)}$ as an indicator of PAH size and $\mathrm{L(7.7\,\mu m)/L(11.3\,\mu m)}$ as an indicator for both ionization and PAH size \citep{smith_07, odowd_09, sales_10, diamond-stanic_10, garciabernete_22}. Both of these ratios are prone to some ambiguity due to a degeneracy between size and ionization, though to a different extent (see D21). In this work, we are primarily interested in studying the competing effects of a possibly changing GSD and changes in the ISRF that heats PAHs. Both of these affect the distribution of power emitted by PAHs at shorter wavelengths versus longer wavelengths. Thus, we specifically consider short-to-long wavelength PAH band ratios in this work, as each of these are necessarily affected by both the GSD and ISRF, regardless of other parameters (such as ionization). 

S07 detected a suppressed $\mathrm{L(7.7\,\mu m)/L(11.3\,\mu m)}$ among LLAGN-hosts compared to normal galaxies in the SINGS sample, and proposed that an LLAGN itself could be directly responsible; in nuclear regions, radiation from the LLAGN could be hard enough to destroy smaller PAH grains while not hard enough to destroy larger grains. These observational results were confirmed by \cite{odowd_09}. \cite{sales_10} and \cite{garciabernete_22} saw similar suppression of both $\mathrm{L(6.2\,\mu m)/L(7.7\,\mu m)}$ and $\mathrm{L(7.7\,\mu m)/L(11.3\,\mu m)}$ in the \textit{Spitzer} spectra for the nuclei of AGN-hosts and noted that their results are consistent with the logic of the selective destruction by AGN hypothesis. Recently, \cite{lai_23} showed using \textit{JWST} spectra that $\mathrm{L(3.3\,\mu m)/L(11.3\,\mu m)}$, a sensitive tracer of PAH size, is suppressed in the nucleus of the Seyfert galaxy NGC\,7469 compared to other regions in the same galaxy. However, in a \textit{Spitzer} study of 35 Seyfert galaxies, \cite{diamond-stanic_10} found significant suppression of $\mathrm{L(7.7\,\mu m)/L(11.3\,\mu m)}$ but not of $\mathrm{L(6.2\,\mu m)/L(7.7\,\mu m)}$, which is another tracer for PAH size. Based on this, they argued against AGN-driven small grain destruction as the cause. 

S07 and \cite{odowd_09} suggested that the usual use of a BPT diagram \citep{baldwin_81} to classify AGN-hosting galaxies may also be leading to the association of AGN with band ratio suppression, since the parameter space occupied by an AGN in a BPT diagram also signals a lack of nuclear star formation which is thought to be the primary source of PAH excitation. In that vein, \cite{kaneda_08} began to address this hypothesis by studying PAH emission in elliptical galaxies with little star formation. Some of these galaxies were normal and some had LLAGN activity. Out of 14 galaxies with significant PAH emission, 12 exhibited suppressed $\mathrm{L(7.7\,\mu m)/L(11.3\,\mu m)}$. This was attributed to a smaller fraction of ionized PAHs due to the UV-poor ISRF of the elliptical galaxies. However, the low average photon energy ($\left<h \nu\right>$) ISRF produced by old stars may contribute to this suppression, and to that of all short-to-long ratios. \cite{ogle} showed that observations of the inner disk of the LLAGN host M58 show an extremely low $\mathrm{L(7.7\,\mu m)/L(11.3\,\mu m)}$ in the inner disk, which they attributed to excitation from old stars emitting relatively soft radiation. Thus, the link between an AGN and low band ratios may be attributable to the relationship between an AGN and the star-formation history of a galaxy, rather than by an impact on the PAH GSD.

Finally, S07 also suggested based on observational evidence that while AGN of sufficient power will destroy PAH grains, LLAGN may be able to supply the UV photons for PAH excitation in the nucleus. Observing the central 10's of pc for AGN-hosts with high angular resolution, \cite{alonso-herrero_14} and \cite{jensen_17} showed that the surface brightness of the 11.3\,$\micron$ feature actually \emph{increased} approaching the nucleus. They argued that these results are consistent with a compact source at the nucleus (which could include the AGN or a nuclear star-forming region) exciting the PAHs. 

It is unclear whether radiation effects or changes in grain properties are responsible for the low short-to-long band ratios observed in LLAGN galaxies. By selecting a target with a radiation field that can be modeled with separable components, and using modeled PAH spectra sensitive to different radiation fields, we may make progress towards quantifying the impact of radiation on band ratios. Selective destruction of small grains by an AGN likely has different impacts at different radii in a galaxy; more small PAHs surviving at radii farther from the nucleus would produce a short-to-long band ratio profile that increases with radius. An observation of this pattern, while accounting for radiation effects with a radiation model, would constitute less ambiguous evidence of AGN influence on PAH properties. Alternatively, the predicted changes in band ratios due to the radiation may be sufficient to produce the observed suppression, rendering small grain destruction unnecessary. 

\section{Sample and Observations} \label{sec:sample}

\begin{figure}
    \includegraphics[width=0.5\textwidth]{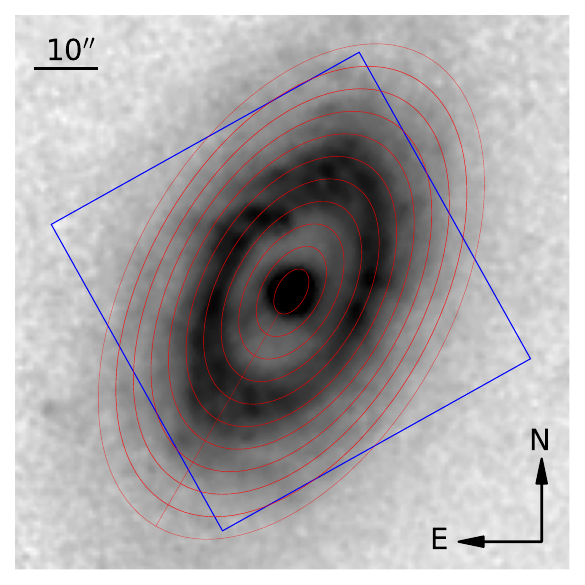}
    \caption{Spitzer IRAC~8\,$\mathrm{\mu m}$ image of the central star-forming ring of NGC\,4138. The footprint of our spectral map is shown as a blue box, and the annular regions we extract spectra from are shown in red.}
    \label{fig:pic}
\end{figure}

Our analysis makes use of \textit{Spitzer}/IRS mapping mode observations of the LLAGN host NGC\,4138 (identified by nuclear X-ray emission via \textit{Chandra} $L_{\mathrm{X}}=10^{41.24}\ \mathrm{erg\,s^{-1}}\ (2-10\,\mathrm{keV})$ \citep{cappi_2006, zhang_2009, Lx_cite}). This galaxy offers a host of benefits for a spatially-resolved study of PAH emission; it is nearby ($D=13.8\,\mathrm{Mpc}$, \cite{sasmirala}), relatively face-on ($\mathrm{inclination=54.1^{\circ}}$ \cite{inclination}), and has an ISRF that can be modeled with simple and separable components (see \S~\ref{sec:radiation model}). Additionally, the IRS map is extensive (see Figure~\ref{fig:pic}). Data from the short (5.2-14.5\,$\micron$) and long (14.0-38.2\,$\micron$) wavelength low resolution modules (SL and LL, respectively) are used, which gives an effective wavelength range of 5.2 -- 38.2\,$\micron$. The SL module produced  a $\sim$1\arcmin $\times$1\arcmin\ square spectral map which is centered on the nucleus of NGC\,4138, and is rotated by $123^{\circ}$ East of North, as shown in Figure~\ref{fig:pic}. Although the LL module produced a considerably larger map, we utilize only the part of the LL map which overlaps the SL map for full spectral coverage. 

\begin{figure}
    \includegraphics[width=0.5\textwidth]{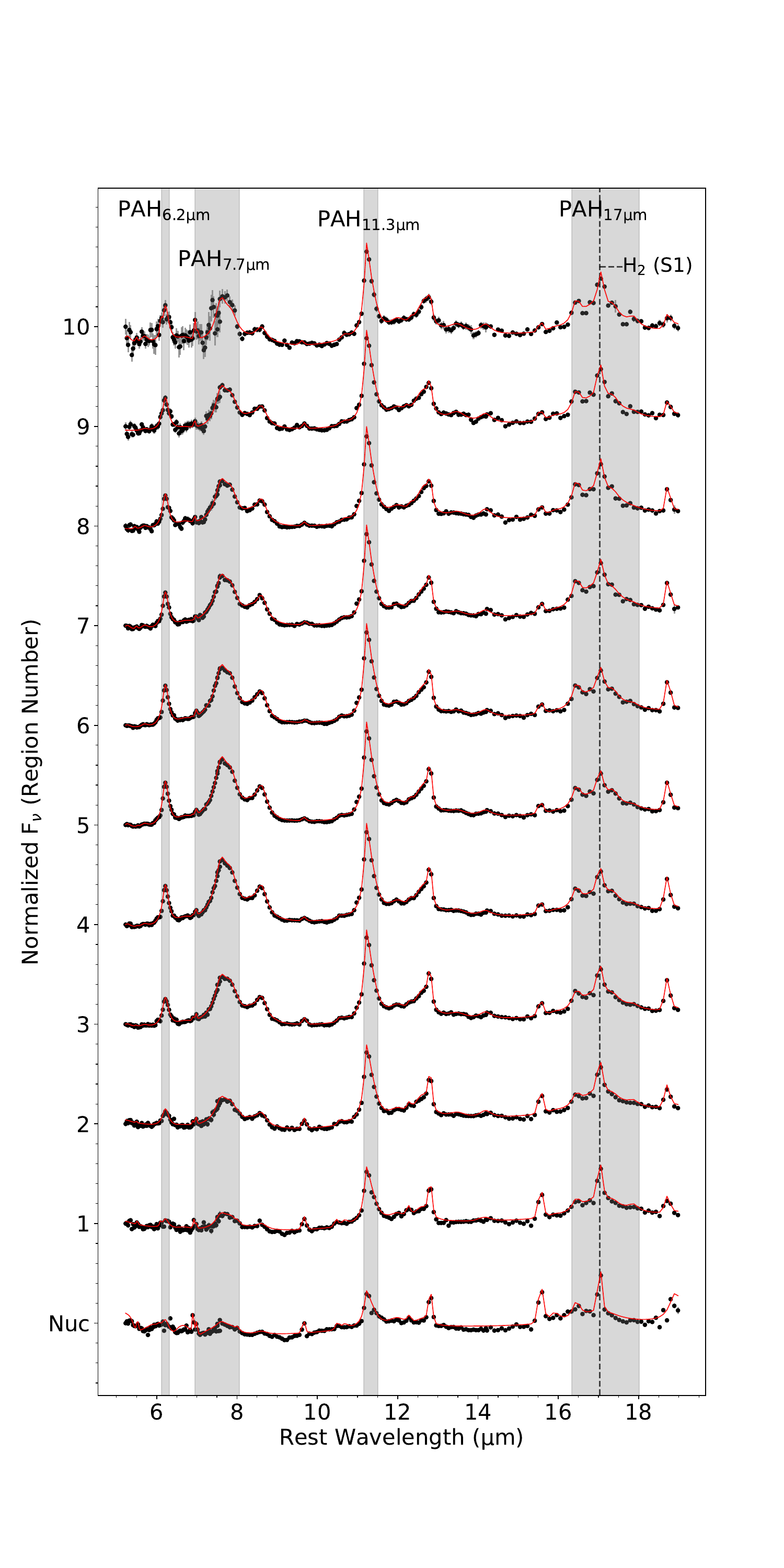}
    \caption{Observed IRS spectrum in each region from Figure~\ref{fig:pic} (black dots). PAHFIT spectra are shown as red curves. Spectra are normalized to the flux at the peak of the 11.3\,$\mathrm{\mu m}$ PAH feature, then offset for clarity. PAH features of interest in this work are shaded in gray. The 17\,$\mathrm{\mu m}$ feature is blended with the H$_2$ 0-0 (S1) line at 17.03\,$\mathrm{\mu m}$, indicated by the vertical dashed line. Labels along the vertical axis indicate the name of region label, with distance from the nucleus increasing upward. Spectra have been truncated for emphasis of PAH features.}
    \label{fig:spectra}
\end{figure}

\begin{figure}
    \includegraphics[width=0.5\textwidth]{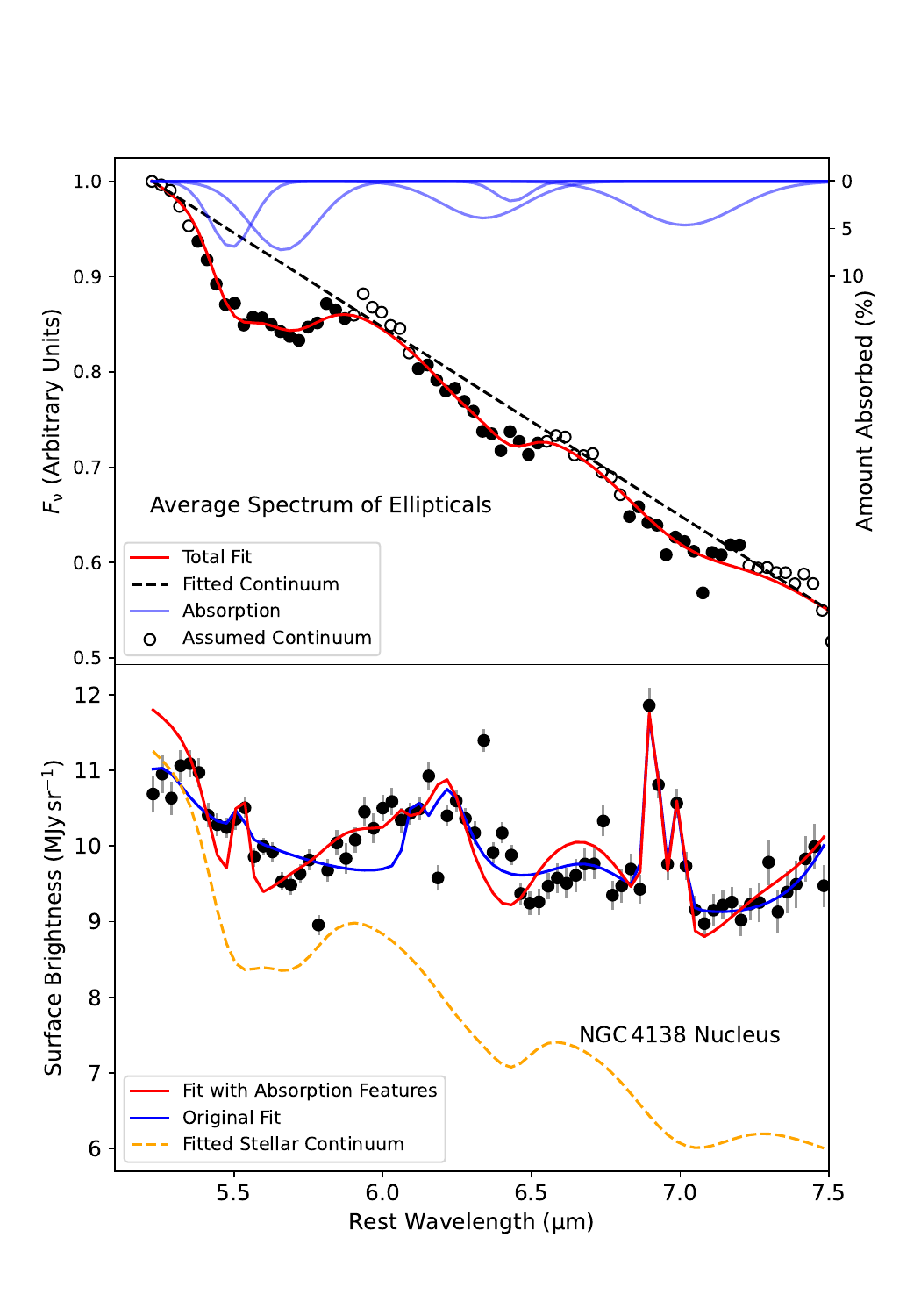}
    \caption{Top: the average elliptical spectrum used to fit the stellar absorption features between 5.5--7.5\,\micron. We assume that the filled circles are within absorption features, while open circles are continuum. Over this small range of wavelengths, we have approximated the continuum as a line. The right vertical axis pertains to the absorption features (blue lines). Bottom: comparison of PAHFIT fits with and without the fitted absorption features in the stellar continuum. Note the broad ``bump" centered at $\sim 6.1\,\micron$.}
    \label{fig:absorption}
\end{figure}

The spectral maps were formatted as spectral cubes using CUBISM \citep{cubism}. We split the galaxy into 10 annular regions which are 4\arcsec\ thick along the major axis (2\farcs3 along minor axis). We also include an elliptical region at the nucleus with a semi-major axis of 4\arcsec (270\,pc at 13.8\,Mpc). The wavelength planes of the cubes are not convolved to a common resolution to preserve the better spatial resolution at short wavelengths. As our goal is to observe the variations in PAH band ratios at small scales, the choice of a 4\arcsec\ thickness is a compromise between maximizing spatial resolution and minimizing aperture effects; the point-spread function (PSF) at the red end of SL has a full-width at half-maximum (FWHM) of $\sim3\farcs9$ \citep{SL_PSF} and the FWHM at the 17\,\micron\ PAH feature is $\sim4\farcs7$. In order to ensure that these small regions do not induce a systematic effect on PAH band ratios, we also test regions which are twice as thick (with the inner-most region unchanged). We do not find a significant difference in PAH band ratios between the two sets of regions (see Figure~\ref{fig:radial_profiles}). All regions are projected onto the plane of the galaxy ($\mathrm{position\ angle=150^{\circ}}$ East of North \citep{hyperleda}) as shown in Figure~\ref{fig:pic}. We extract spectra from these regions, and a scaling correction is applied to the SL spectra such that the median flux of the five longest wavelength SL data points matches the median flux of the five shortest wavelength data points of the LL spectrum. Normalized spectra for each region are shown in Figure~\ref{fig:spectra}. Spectral decomposition in each region is performed using PAHFIT (S07), and the fitted spectra are shown as red lines in Figure~\ref{fig:spectra}. PAHFIT allows us to obtain the continuum-subtracted integrated luminosity of individual PAH features.

The spectrum of our nuclear region exhibits an apparent emission feature from $\sim5.7 - 6.5\,\micron$ that is not present in the spectra extracted from our other regions. This feature interferes with the extraction of PAH fluxes. We attribute this to absorption features in the starlight produced by the evolved stellar population of the bulge; a similar apparent feature was noted in the spectra of stars in Milky Way globular clusters by \cite{Sloan2010}, who also suggested absorption as the cause. To account for this, we searched for archival \textit{Spitzer}/IRS data of elliptical galaxies which show these absorption features in their spectra; we assume that these serve as an appropriate analog for the bulge of NGC\,4138. We found five objects that show the absorption features: NGC\,821, NGC\,1549, NGC\,3904, NGC\,4621, and NGC\,4660, and averaged the IRS spectra at their centers ($r<1\farcs5$).

In addition to the absorption at $\sim5.7$ and $6.5\,\micron$, there is an additional feature at $7.0\,\micron$. We find that five Gaussian-shaped features adequately describe the observed absorption when a linear continuum is assumed. In PAHFIT, we apply these absorption features at fixed ratios to the standard assumed stellar modified black-body ($T=5000\,\mathrm{K}$), with the strength of the features as a free parameter. See Figure~\ref{fig:absorption} for the fit to the average elliptical galaxy spectrum and a zoom-in at the wavelengths of interest for our inner-most region. 

\section{PAH Emission Model} \label{sec:emission}

The D21 models predict PAH spectra for various sizes of PAH grains and are computed for 14 different incident starlight spectra with different population ages and parameters. In addition, each spectrum includes an extinguished version through a slab of dust with total $A_{\mathrm{V}}=2\,\mathrm{mag}$, and each spectrum is varied across a range of intensity. We refer to the combination of the mix of starlight spectra and the intensity as an ISRF. Finally, the D21 model PAH spectra are provided for neutral and singly ionized grains. In this work, we use PAHFIT to obtain the band ratios of the modeled PAH spectra and compare to those of the observed spectra.

Real PAH populations in galaxies exist in a mix of grain sizes and ionization states. Therefore to produce realistic model PAH spectra we average the contributing spectra of many different model PAH grains, using an assumed distribution of PAH size and ionization as the weighting function. For PAH size, we adopt the double log-normal grain size distribution (GSD) described in D21:

\begin{equation} \label{GSD}
    \frac{1}{n_{\mathrm{H}}} \frac{d n_{\mathrm{PAH}}}{d a}=\sum_{j=1}^2 \frac{B_j}{a} \exp \left\{-\frac{\left[\ln \left(a / a_{0 j}\right)\right]^2}{2 \sigma^2}\right\}
\end{equation}
where $a$ is the radius of a PAH grain. Each $B_j$, $a_{0j}$, and $\sigma$ sets the amplitude, horizontal position, and width, respectively. In this work, we express PAH size as the number of carbon atoms per molecule $N_{\mathrm{C}}$, and we adopt the relationship between $a$ and $N_{\mathrm{C}}$ in D21:

\begin{equation} \label{a_to_Nc}
N_{\mathrm{C}}=418\left(\frac{a}{10\,\mathrm{\AA}}\right)^3\ .
\end{equation}

\begin{figure}
    \includegraphics[width=0.5\textwidth]{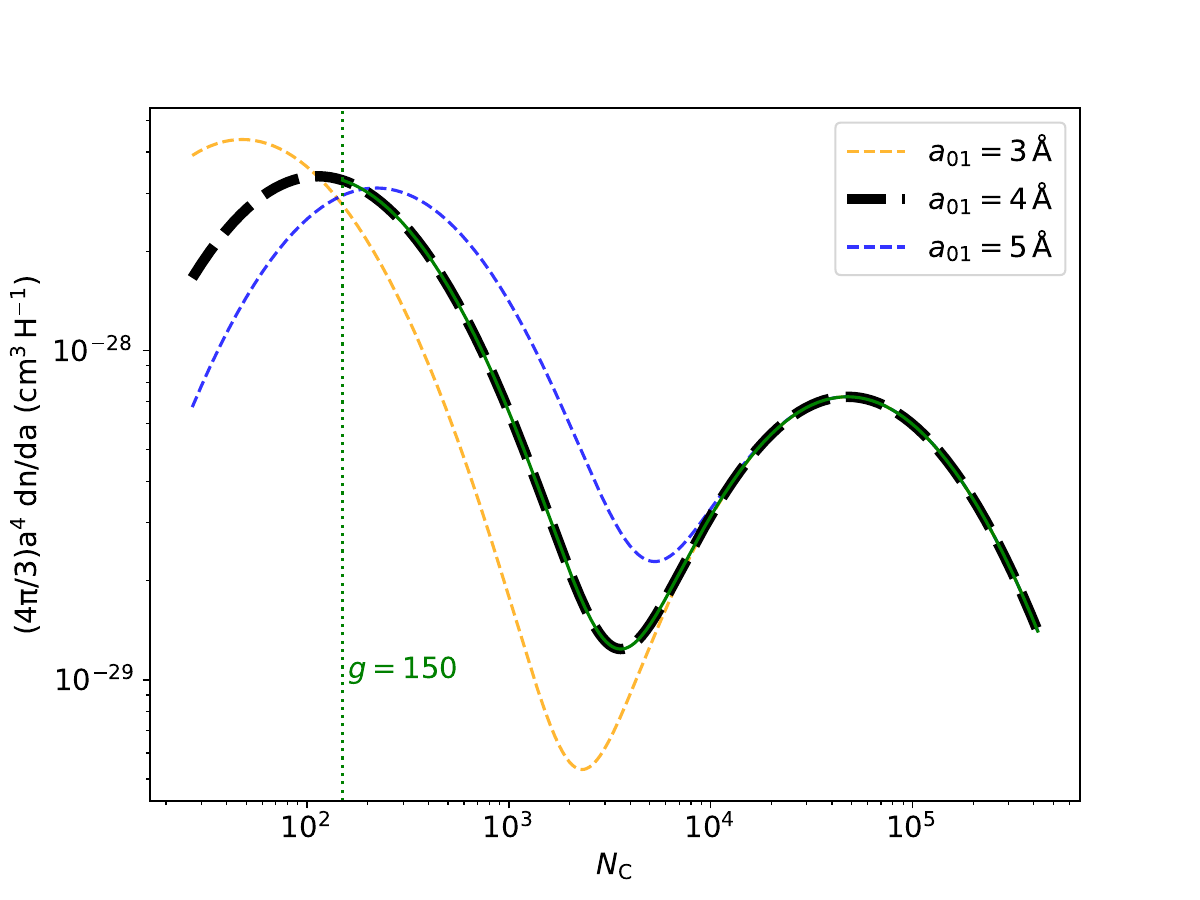}
    \caption{Example PAH grain size distributions. (See Eq.~\ref{GSD}). The three examples given show the small, standard, and large distributions from D21 as the dashed yellow, black, and blue curves, respectively. The thick black curve emphasizes that we assume NGC\,4138 to have a standard distribution. The green curve serves as an example of a standard GSD with a lower size limit of $g=150$.}
    \label{fig:gsd}
\end{figure}

Following D21, we always fix $a_{02}$ and $B_2$ at $30.0\,\mathrm{\AA}$ and $3.113\times 10^{-10}\,\mathrm{H}^{-1}$, respectively, and $\sigma=0.40$. GSDs with a larger or smaller average PAH size are produced by changing $a_{\mathrm{01}}$, which shifts the log-normal for the smallest PAHs, as shown in Figure~\ref{fig:gsd}. For a given $a_{\mathrm{01}}$, we vary $B_1$ such that the volume of the PAHs in this log-normal is always $V_{\mathrm{PAH,1}}=3\times10^{-28}\,\mathrm{cm^{3}\,H^{-1}}$, as in D21. Figure~\ref{fig:gsd} shows the three GSDs discussed in D21, called ``small" ($a_{\mathrm{01}}=3\,\mathrm{\AA}$), ``standard" ($a_{\mathrm{01}}=4\,\mathrm{\AA}$), and ``large" ($a_{\mathrm{01}}=5\,\mathrm{\AA}$). In this work, we assume NGC\,4138 has a ``standard" GSD.

We wish to examine the effect that an AGN may have on the GSD, and in particular we consider the prevalent hypothesis that an AGN selectively destroys small PAH grains. To simulate this, we introduce a parameter $g$ which is the minimum surviving PAH size (measured in $N_{\mathrm{C}}$) for a given PAH population. We do not re-scale the GSD to preserve the PAH volume when a larger $g$ is imposed. This is consistent with small grain destruction. The green curve in Figure~\ref{fig:gsd} shows the ``standard" GSD with an example lower PAH size limit at $g=150$. In D21, $g=27$ is used for all considered GSDs.

\begin{figure}
    \includegraphics[width=0.5\textwidth]{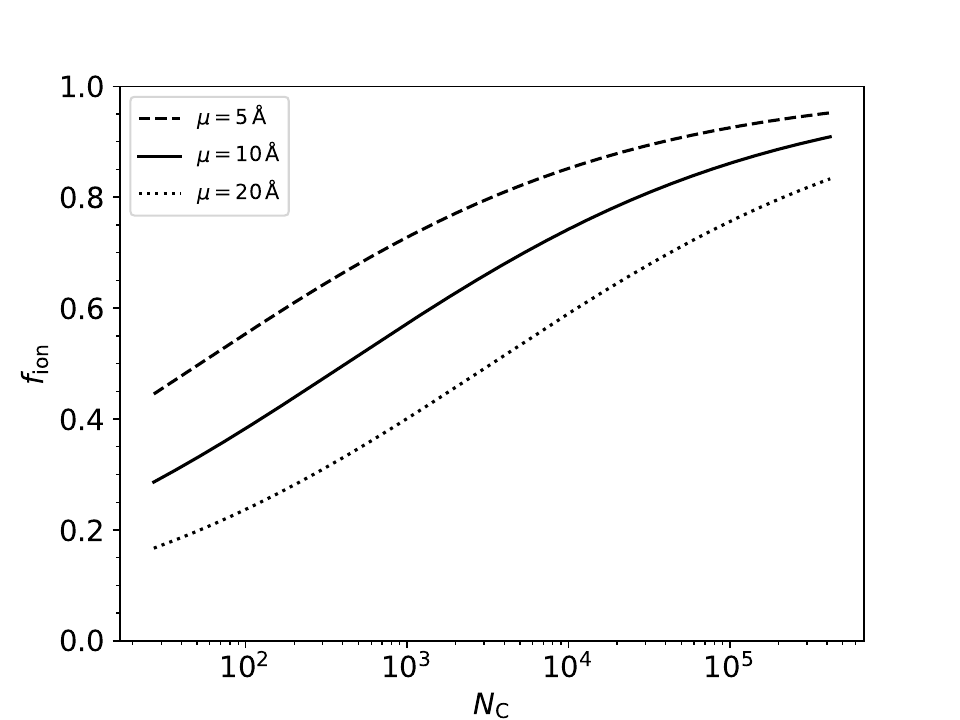}
    \caption{PAH ionization fraction distribution described by Eq.~\ref{fion} with the three values discussed in D21. The dotted line represents ``low" ionization, the solid line represents ``standard" ionization, and the dashed line is ``high" in D21.}
    \label{fig:ifd}
\end{figure}

We also adopt the functional form used by D21 for PAH ionization, in which the fraction of PAH grains that are singly ionized is a function of the PAH grain size: 
\vspace*{-1mm}
\begin{equation} \label{fion}
f_{\mathrm{ion}}(a)=1-\frac{1}{ 1 + \frac{a}{\mu}} 
\end{equation}
for a PAH of radius $a$ in a distribution where 50\% of PAH grains with $a = \mu$ are ionized (see Figure~\ref{fig:ifd}). Thus, we vary $\mu$ when we wish to vary the ionization fraction distribution.

D21 uses the dimensionless parameter $U$ to denote the intensity of starlight that PAHs are exposed to; it is the dust heating rate due to starlight relative to that of the ISRF near the Sun \citep{MMP}. The $U$ and the fractional contribution of each incident starlight spectrum at each radius is determined by our radiation model of NGC\,4138, as discussed in the next section. We do not apply any scaling to model PAH spectra, since we are only interested in the PAH band ratios.

\section{Radiation Model of NGC\,4138} \label{sec:radiation model}

The ISRF in a galaxy often includes contributions from a complicated mixture of different stellar populations and other luminous objects --- such as an AGN --- all of which may contribute to PAH excitation. Adding additional complexity, this radiation blend varies across the different regions of a galaxy. In order to investigate the effects of varying ISRF radiation intensity and hardness on PAH emission, the contributions of competing radiation sources must be separated and modeled independently. NGC\,4138 provides a convenient environment for this, with a pronounced star-forming ring embedded in a smooth disk of older stars \citep{ngc4138_disk_properties, ngc4138_ages}. In this section, we describe our radiation model of NGC\,4138 which serves three major purposes: 1) to predict the overall radiation intensity experienced by PAHs at different radii, 2) to predict the mix of starlight spectra experienced by PAHs at different radii, and 3) to investigate the possibility of PAH excitation by an AGN. 


\subsection{Bulge-Disk-Ring Decomposition} \label{decomp}

\begin{table*}
\caption{Summary of Galaxy Component Decomposition}
\centering
\begin{tabular}{cccc}
\hline
\textbf{Component} & \textbf{Used Bandpass}& \textbf{$I_{\mathrm{0,comp}}$}& \textbf{Spatial Parameters} \\
\hline
Bulge & IRAC 3.6\,\micron &  $13.54\,\mathrm{MJy\,sr^{-1}}$ & $R_{\mathrm{bulge}}=187\,\mathrm{pc}$, $n=1.57$\\
Stellar Disk & IRAC 3.6\,\micron & $7.68\,\mathrm{MJy\,sr^{-1}}$ & $R_{\mathrm{disk}}=992\,\mathrm{pc}$\\
Ring & IRAC 8\,\micron & $3.99\,\mathrm{MJy\,sr^{-1}}$ & $R_{\mathrm{ring}}=1.2\,\mathrm{kpc}$, $\sigma_{\mathrm{ring}}=0.30\,\mathrm{kpc}$\\
\hline
\label{tab:decomp}
\end{tabular} 

Compare values to those from \cite{fisher_and_drory}: $R_{\mathrm{bulge}}=182\,\mathrm{pc}$, $n=1.65$, $R_{\mathrm{disk}}=1.17\,\mathrm{kpc}$. 

The scale length of the exponential component in the IRAC 8\,$\mathrm{\mu m}$ filter is $R_{\mathrm{disk,8\,\micron}}=3.66\,\mathrm{kpc}$.
\end{table*}

Our approach to creating a radiation model is similar to that of \cite{andromedas_dust}, who demonstrated how dust properties in M31 vary with the changing intensity of the radiation environment. This was done by separately modeling the PAH heating due to the bulge and disk of M31. In our radiation model, NGC\,4138 is idealized as four components: an AGN, a bulge, a stellar disk, and a star-forming ring embedded in the disk. Separating these is crucial for testing how each individually may influence PAH excitation. We fit luminosity distributions (via surface brightness profiles) for the bulge, stellar disk, and ring. The fitted parameters of these profiles describe the spatial distributions of the luminosity source functions for each component in the radiation model. Note that the regions we sample photometry from are different between photometric filters, and are also not the same as the regions we extract spectra from. This is done to take advantage of the varying spatial resolutions between filters. 

\begin{figure}
    \includegraphics[width=0.5\textwidth]{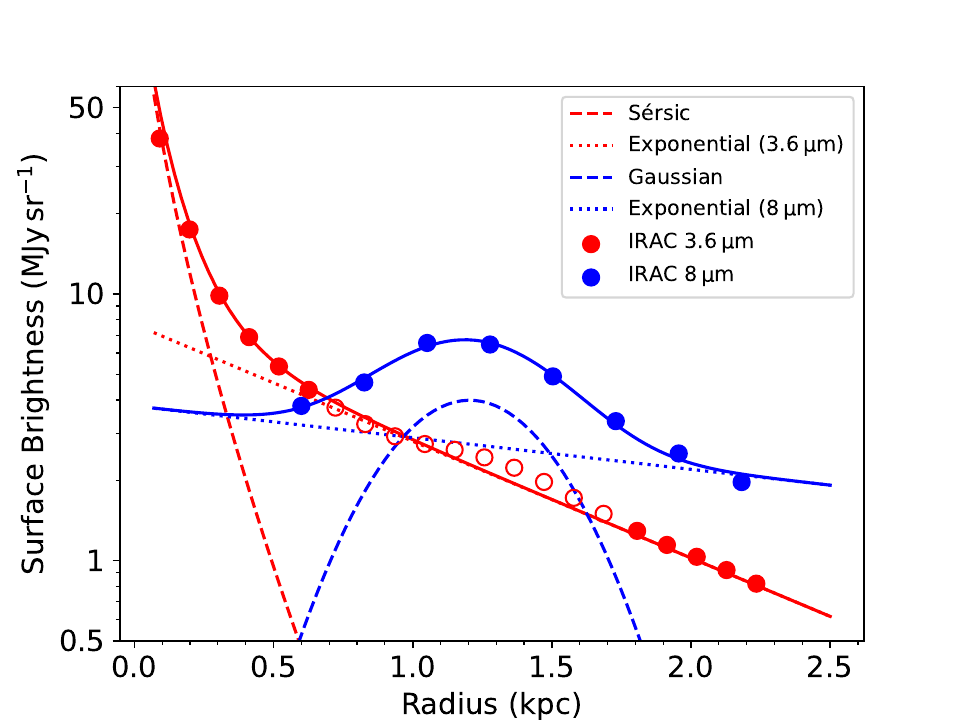}
    \caption{Surface brightnesses observed at annuli of varying radius projected onto the plane of NGC\,4138. Light from old stars is traced by IRAC~3.6\,\micron\ photometry as red dots, and PAH light is traced by the IRAC~8\,\micron\ photometry as blue dots. The solid red curve is the fit to the 3.6\,\micron\ photometry, as the sum of an exponential (red dotted line) and a S\'ersic profile (red dashed line). The solid blue curve is fitted to the 8\,\micron\ photometry, as the sum of an exponential (dotted blue line) and a Gaussian profile (dashed blue line), which we use to describe the radial shape of ring luminosity. Open red circles within ring-dominated radii are observed 3.6\,\micron\ photometry that were not used in the fit (see \S~\ref{sec:radiation model}). Typical uncertainties are $\sim 7.6 \times 10^{-4}$ and $\sim 3.8 \times 10^{-3}\,\mathrm{MJy\,sr^{-1}}$ for the 3.6\,\micron\ and 8\,\micron\ photometry, respectively.}
    \label{fig:bulgefit}
\end{figure}

We sample projected annular regions on a \textit{Spitzer}/IRAC 3.6\,$\mathrm{\mu m}$ image to obtain a surface brightness radial profile for the bulge and stellar disk, since emission in this filter is dominated by light from old stellar populations \citep{meidt_2012}. The inner diameter of the inner-most annulus was chosen to be larger than the PSF FWHM of IRAC 3.6\,$\mathrm{\mu m}$ (1\farcs6), to avoid contamination from the AGN. Subsequent annular regions are each 1\farcs6 ($\sim$100\,pc) thick. In addition, we do not include measurements of radius between 10--27\,\arcsec ($\sim$670--1800\,pc) in the fit, because the star-forming ring contributes noticeably to the 3.6\,$\mathrm{\mu m}$ surface brightness between these radii. We then fit the surface brightness profile as a linear combination of a S\'ersic profile \citep{sersic} which describes the bulge, 

\begin{equation} \label{I_bulge}
I_{\mathrm{bulge}}(r)=I_{\mathrm{0,bulge}} \exp \left\{-b_n\left[\left(\frac{r}{R_\mathrm{bulge}}\right)^{1 / n}-1\right]\right\}
\end{equation}

\noindent
where $R_{\mathrm{bulge}}$ is the half-light radius of the bulge, $I_{\mathrm{0,bulge}}$ is the surface brightness at $R_{\mathrm{bulge}}$, and $n$ is the S\'ersic index. Similar to \cite{fisher_and_drory}, we use the approximation $b_n\approx 2.17n-0.355$, which applies for $n>0.33$. The exponential profile describes the stellar disk of the galaxy \citep{exponential_disk},

\begin{equation} \label{I_disk}
I_{\mathrm{disk}}(r)=I_{\mathrm{0,disk}} \exp \left\{-\frac{r}{R_{\mathrm{disk}}}\right\}
\end{equation}

\noindent
where $R_{\mathrm{disk}}$ is the scale length of the stellar disk, and $I_{\mathrm{0,disk}}$ is the central surface brightness of the disk. The surface brightness profile, the fitted S\'ersic, and the fitted exponential profile are shown in Figure~\ref{fig:bulgefit}. The fitted $R_{\mathrm{bulge}}$, $n$, and $R_{\mathrm{disk}}$ are listed in Table~\ref{tab:decomp}.

We follow a similar procedure to describe the surface brightness profile of the star-forming ring of the galaxy. We sample projected annular regions of a \textit{Spitzer}/IRAC 8\,$\mu m$ image, which traces primarily PAH emission. In this case, projected annular regions are 1\farcs98 ($\sim$130\,pc) thick, since this is the PSF FWHM of IRAC \,8\micron. We fit this profile as the linear combination of an exponential profile (in the form of Eq. \ref{I_disk}) for the stellar disk of the galaxy and a Gaussian profile, which we find suitably describes the star-forming ring:

\begin{equation} \label{I_ring}
I_{\mathrm{ring}}(r)=I_{\mathrm{0,ring}} \exp \left\{-\frac{ (r-R_{\mathrm{ring}})^2 }{2 \sigma_{\mathrm{ring}}^2}\right\}
\end{equation}

\noindent
where $R_{\mathrm{ring}}$ is the radius at which the surface brightness of the ring is greatest, $\sigma_{\mathrm{ring}}$ is a measure of the ring thickness, and $I_{\mathrm{0,ring}}$ is the surface brightness of the ring at $R_{\mathrm{ring}}$. See Figure~\ref{fig:bulgefit} for the observed and fitted surface brightness profile of the ring, and Table~\ref{tab:decomp} for fitted parameters. 

\subsection{Scaling the Luminosity Source Functions} \label{sec:scaling}

In this work, we assume that energies between 0--13.6\,eV can excite PAHs, since the D21 models are calculated for PAHs within regions where hydrogen gas is not photoionized. For each component, we choose a template spectrum to be scaled and integrated over for obtaining the luminosity in this energy range. Though the expression defining the source function for each galaxy component is different, each requires a scale factor. What follows is our method for determining these scale factors.

\textbf{AGN:} The D21 models do not include an AGN spectrum as one of the PAH-heating ISRFs. Thus, the AGN is assumed to have the spectrum of the Seyfert\,2 composite template from \cite{agn_template} for the purposes of luminosity scaling. We extend this spectrum with a modified black-body between 1000 and 8000\,\micron\ ($T=100\,K$) so that it has the same range of wavelengths as the assumed spectra of the other galaxy components. The template spectrum is scaled such that the integrated $\mathrm{2-10\,keV}$ X-ray luminosity of the template spectrum is equal to the observed intrinsic absorption-corrected value for the nucleus of NGC\,4138, $L_{\mathrm{X}}=10^{41.24}\,\mathrm{erg\,s^{-1}}$ \citep[2--10\,keV,][]{Lx_cite}. Upon integrating the scaled spectrum from 0--13.6$\,\mathrm{eV}$, we obtain the scale factor $L_{\mathrm{AGN}}$. For the purposes of model PAH spectra, we choose the 3\,Myr stellar spectrum from \cite{bc03} to represent PAHs irradiated by an AGN, since it is the highest $\left<h \nu\right>$ spectrum included in the D21 models.

\textbf{Bulge:} The function in our radiation model that describes the distribution of luminosity due to the bulge is the deprojection of the S\'ersic profile proposed by \cite{PG_97} (see \S~\ref{sec: energy density curves} and Eq.~\ref{L_bulge}). Since we are interested in a distribution of luminosity rather than mass, we assume a mass-to-light ratio of unity and this function is scaled by

\begin{equation} \label{3Dsersic}
\rho_{\mathrm{bulge}} = \Sigma_{\mathrm{0,bulge}} \mathrm{e}^{b_{n}} b^{n(1-p_{n})}_{n} \frac{\Gamma(2 n)}{2 R_{\mathrm{bulge}} \Gamma(n(3-p_{n}))} \text{ ,}
\end{equation}

\noindent where $\Sigma_{\mathrm{0,bulge}}$ is the luminosity surface density at $R_{\mathrm{bulge}}$, $n$ and $b_n$ are the same as in Table~\ref{tab:decomp}, and we use the calibration from \cite{LimaNeto1999} that

\begin{equation} \label{p_equation}
p_n \approx 1 - 0.6097/n + 0.05463/n^2 \text{ .}
\end{equation}

\noindent We choose the M31 bulge spectrum from \cite{groves_12} to represent the bulge spectrum of NGC\,4138 in our radiation model, since it is included among the D21 ISRFs. This is scaled such that $I_{\mathrm{0,bulge}}$ is recovered when the spectrum is passed through the IRAC 3.6\,$\mathrm{\mu m}$ filter. This scaled spectrum is then integrated from 0--13.6$\,\mathrm{eV}$ to get $\Sigma_{\mathrm{0,bulge}}$.

\textbf{Stellar Disk:} The stellar disk of old stars is also assumed to have the M31 bulge spectrum from \cite{groves_12}, as the color of the disk at radii past the ring is similar to that of the bulge \citep{disk_color}. This spectrum is scaled to recover $L_{\mathrm{disk,3.6\mu m}}$ when passed through the IRAC 3.6\,$\mu m$ filter, where $L_{\mathrm{disk,3.6\mu m}}$ is the luminosity contained within the aperture corresponding to $I_{\mathrm{0,disk}}$. We find $L_{\mathrm{disk,3.6\mu m}}=8.23\times10^{7}\,\mathrm{L_{\odot}}$. The scaled spectrum is integrated from 0--13.6$\,\mathrm{eV}$ and divided by the area of the same aperture to obtain the scale factor $\Sigma_{\mathrm{disk}}$.

\textbf{Star-Forming Ring:} 
Following the method described in \cite{galex_mips}, we obtain a star formation rate (SFR) \citep{SFR2} from a combination of GALEX far-ultraviolet (FUV) and \textit{Spitzer}/MIPS 24\,$\,\mathrm{\mu m}$ images. We find $\mathrm{SFR=0.20\,M_{\odot}\,yr^{-1}}$. All photometry (GALEX FUV and MIPS) uses a projected annular aperture with an inner radius of 0.57\,kpc and an outer radius of 2.1\,kpc. This inner radius is just large enough to exclude the AGN contribution to the \textit{Spitzer}/MIPS 24\,$\,\mathrm{\mu m}$ image (PSF FWHM = 5\farcs9) and the outer radius is 3\,$\mathrm{\sigma_{\mathrm{ring}}}$ from $R_{\mathrm{ring}}$. We then use the GALEX FUV-SFR relation from \cite{salim_07} to obtain an attenuation-corrected luminosity of the ring in the GALEX FUV bandpass, $L_{\mathrm{ring,FUV}}=9.60\times10^{8}\,\mathrm{L_{\odot}}$. The star formation contained within the ring is assumed to produce the spectrum of the 10\,Myr stellar population from \cite{bc03}, which is also included as a D21 ISRF. This spectrum is scaled such that $L_{\mathrm{ring,FUV}}$ is recovered when the spectrum is passed through the GALEX FUV bandpass. Thus, the scale factor $\Sigma_{\mathrm{ring}}$ is obtained by integrating the scaled spectrum from 0--13.6$\,\mathrm{eV}$ and dividing by the area of the aperture used to get the GALEX SFR.

\subsection{Component-wise Energy Density Curves} \label{sec: energy density curves}

\begin{table*}
\caption{Summary of Radiation Model Source Functions}
\centering
\begin{tabular}{ccccc}
\hline
\textbf{Component} & \textbf{Source Function Shape}& \textbf{Scale Factor}& \textbf{Total 0--13.6\,eV Luminosity}& \textbf{ISRF Spectrum} \\
\hline

AGN & Point-like (Eq. \ref{L_AGN}) & $L_{\mathrm{AGN,0-13.6\,eV}}=9.718\times10^{8}\,\mathrm{L_{\odot}}$ & $9.718\times10^{8}\,\mathrm{L_{\odot}}$ & 3\,Myr \\

Bulge & 3-D S\'ersic (Eq. \ref{L_bulge}) & $\rho_{\mathrm{bulge,0-13.6\,eV}}=164.0\,\mathrm{L_{\odot}\,pc^{-3}}$ & $8.356\times10^{9}\,\mathrm{L_{\odot}}$ & M31 Bulge \\

Stellar Disk & 2-D Exponential (Eq. \ref{L_disk}) & $\Sigma_{\mathrm{disk,0-13.6\,eV}}=1329\,\mathrm{L_{\odot}\,pc^{-2}}$ & $6.936\times10^{9}\,\mathrm{L_{\odot}}$ & M31 Bulge \\

Ring & 2-D Gaussian (Eq. \ref{L_ring}) & $\Sigma_{\mathrm{ring,0-13.6\,eV}}=710.3\,\mathrm{L_{\odot}\,pc^{-2}}$ & $4.035\times10^{9}\,\mathrm{L_{\odot}}$ & 10\,Myr\\

\hline
\label{tab:ucurve}
\end{tabular}
\end{table*}

\begin{figure*}
    \includegraphics[width=1\textwidth]{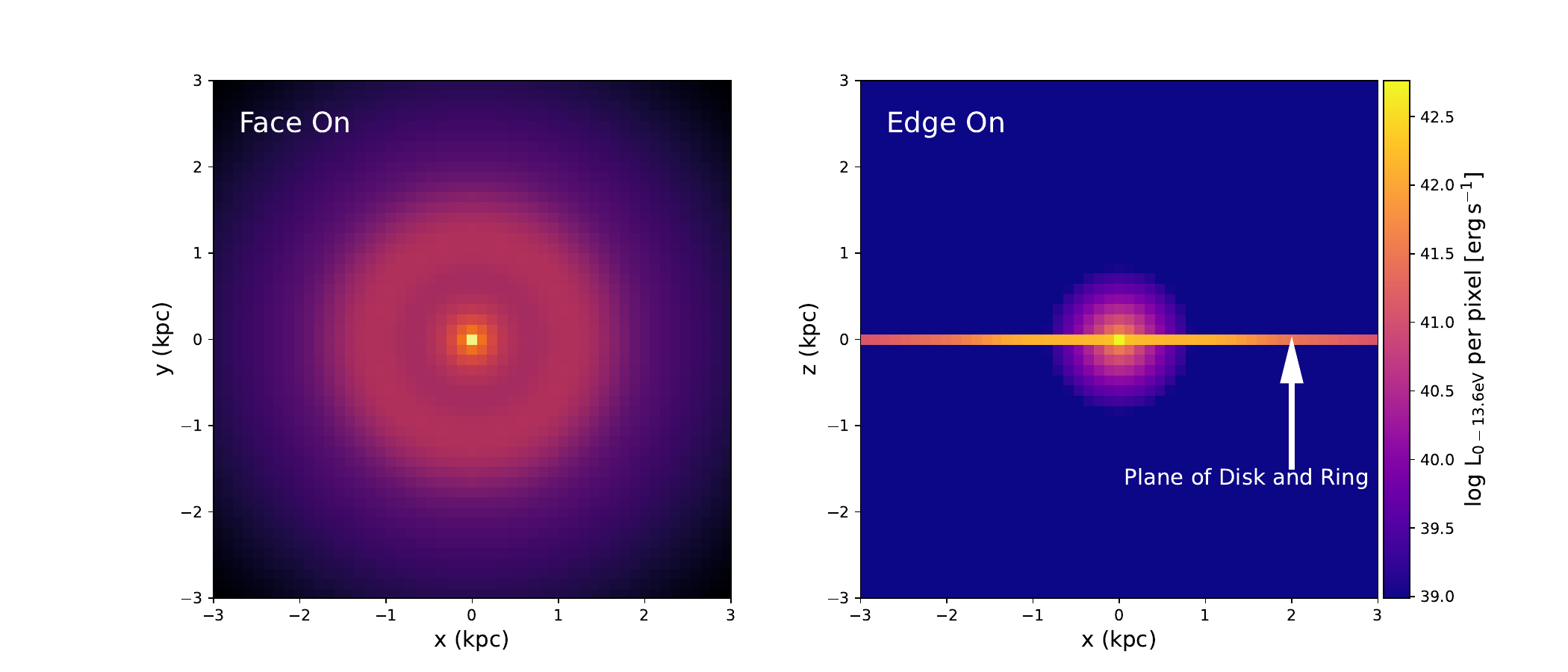}
    \caption{The radiation model summed along the $z$ and $y$ directions are shown in the left and right panels, respectively. Colors indicate the combined luminosity integrated from 0--13.6\,eV from each galaxy component in a given voxel. Here, we show a lower resolution $51\times51\times51$ voxel version of the model to accentuate plane of the disk and ring in the edge on view. This lower resolution is for visualization purposes only.}
    \label{fig:modelpic}
\end{figure*}

The purpose of the radiation model is to predict the energy density contributed to a region by the ISRF of each galaxy component, in order to know what kind of radiation PAHs are being exposed to. To facilitate this, we construct a 3-dimensional array of volumetric pixels (voxels) which contains the radiation model. The array has dimensions $501\times501\times501$ ($x, y, z$, origin at center), mapping to 6\,kpc on a side. Thus, a single voxel has an edge length of $\sim$12\,pc. For each of the four galaxy components, voxels are filled with a luminosity density according to their respective source functions. The AGN occupies the single voxel at the center of the cube,

\begin{equation} \label{L_AGN}
\mathcal{L}_{\mathrm{AGN,0-13.6\,eV}}(x,y,z)=L_{\mathrm{AGN}}, \text{ at }x,y,z=0 \text{ .}
\end{equation}

\cite{ngc4138_disk_properties} obtained rotation curves for NGC\,4138 by assuming a zero-thickness mass distribution for the stellar disk and a spherical mass distribution for the bulge. We make the same assumption, although we are distributing luminosity sources rather than mass in our model. Thus, the bulge is the single galaxy component that we model as 3-dimensional, and we fill voxels with bulge luminosity according to the approximate deprojected 3D S\'ersic profile from \cite{PG_97}


\begin{equation} \label{L_bulge}
\begin{aligned}
\mathcal{L}_{\mathrm{bulge,0-13.6\,eV}}(x,y,z)= \\
& \hspace*{-1.3in} \rho_{\mathrm{bulge}} \left(\frac{\sqrt{x^2+y^2+z^2}}{R_{\mathrm{bulge}}}\right)^{-p_n} \\
& \exp \left\{-b_n\left[\left(\frac{\sqrt{x^2+y^2+z^2}}{R_{\mathrm{bulge}}}\right)^{1 / n}-1\right]\right\}
\text{ ,}
\end{aligned}
\end{equation} 

\noindent and voxels are filled with disk luminosity (in the $x$-$y$ plane) via 

\begin{equation} \label{L_disk}
\mathcal{L}_{\mathrm{disk,0-13.6\,eV}}(x,y,z)=\Sigma_{\mathrm{disk}} \exp \left\{-\frac{\sqrt{x^2+y^2}}{R_{\mathrm{disk}}}\right\}
\text{, at $z=0$ .}
\end{equation}

\noindent The ring is also confined to the $x$-$y$ plane, and it fills voxels as the Gaussian

\begin{equation} \label{L_ring}
\begin{aligned}
\mathcal{L}_{\mathrm{ring,0-13.6\,eV}}(x,y,z)=\\
& \hspace*{-1.3in} \Sigma_{\mathrm{ring}} \exp \left\{-\frac{ (\sqrt{x^2+y^2}-R_{\mathrm{ring}})^2 }{2 \sigma_{\mathrm{ring}}^2}\right\}
\text{, at $z=0$ .}
\end{aligned}
\end{equation}

\noindent Note that we do not include the exponential component in the ring source function to avoid double-counting the stellar disk. Additionally, Eq.~\ref{L_bulge} yields infinity at the center voxel, so we smooth this to the average of the bulge-values of adjacent voxels.

The energy density contributed by a galaxy component within a given voxel, $u_{\mathrm{comp,0-13.6\,eV}}(x,y,z)$ (``comp" = component), can be expressed as 

\begin{equation} \label{u_curve}
\begin{aligned}
u_{\mathrm{comp,0-13.6\,eV}}(x,y,z)= \\
& \hspace*{-1in} \frac{1}{4\pi c} 
\iint_V \frac{\mathcal{L}_{\mathrm{comp,0-13.6\,eV}}(x',y',z')dx' dy' dz'}{(x'-x)^2+(y'-y)^2+(z'-z)^2}
\text{ .}
\end{aligned}
\end{equation}

\noindent We make the assumption that PAHs are primarily heated by the diffuse radiation in the ISM \citep{aniano_2012, aniano_2020}, so for each test location we exclude contributions from voxels within 15\,pc, just slightly larger than our chosen voxel edge-length. This also avoids infinities when computing $u_{\mathrm{comp}}$. In \S~\ref{sec:discussion}, we discuss the effect that PAHs heated by nearby photodissociation regions might have on our radiation model. See Figure~\ref{fig:modelpic} for a visualization of the radiation model. Here, it is shown at a lower resolution so that the thin plane of the ring+stellar disk can be seen.

The attenuating effects of dust influence the way that radiation is able to propagate throughout the ISM. In NGC\,4138, the dust is concentrated in the star-forming ring (see Figure~\ref{fig:pic}), and we expect this to have two main consequences on the exchange of radiation between galaxy components. First, the radiation emitted by the AGN would be truncated at the inner radius of the ring upon encountering the higher concentration of dust. Second, much of the radiation that originates from the star-formation within the ring would not escape it before being reddended or absorbed. A full incorporation of this radiative transfer is outside of the scope of this work, but we do consider a high-extinction bounding case where AGN radiation is truncated at the inner radius of the ring, and radiation originating from the ring is truncated at the edges of the ring. We define the edges of the ring as the same ring region as was used to obtain $L_{\mathrm{ring,FUV}}$ (see \S~\ref{sec:scaling}). 

\begin{figure*}
    \centering
    \includegraphics[width=1\textwidth]{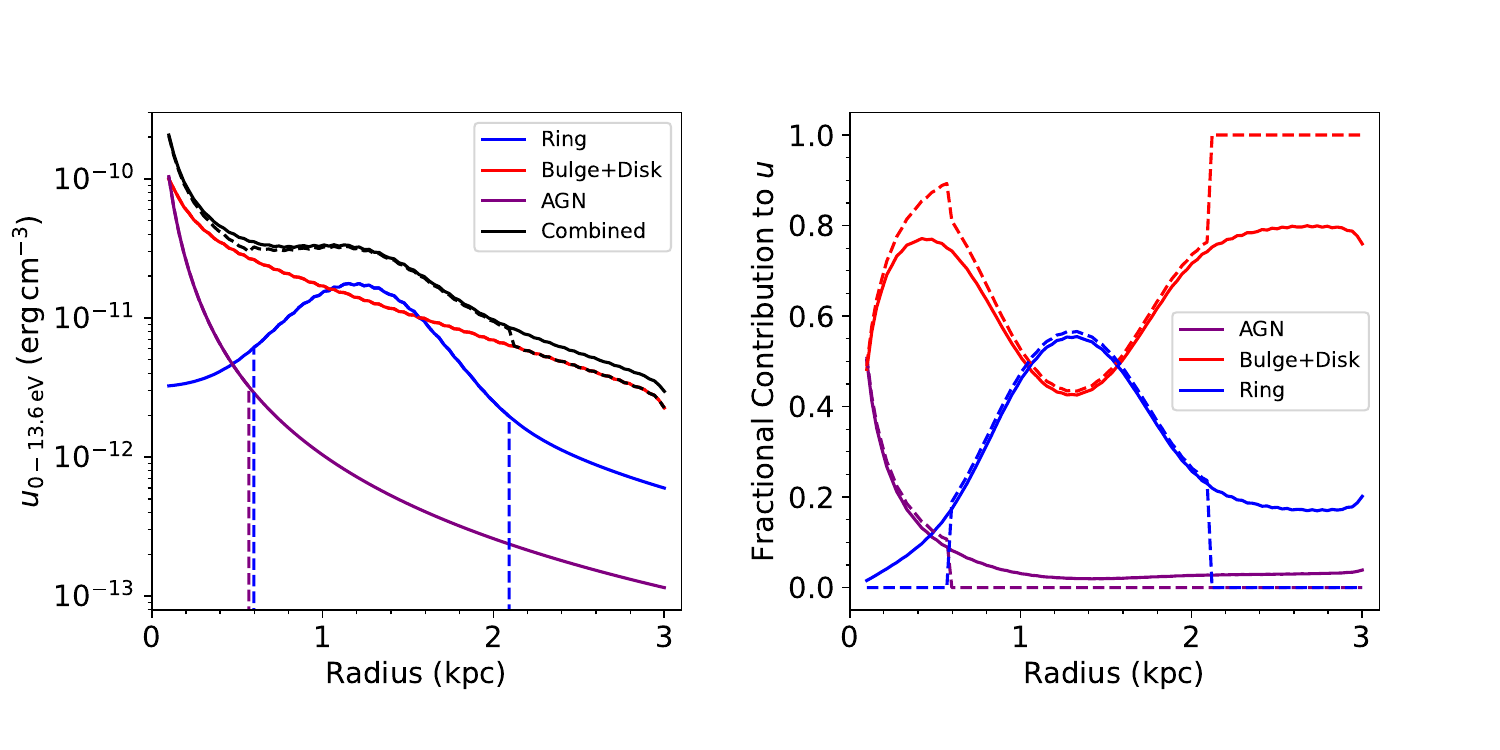}
    \caption{Left panel: colored lines describe the energy density $u$ integrated between 0 -- 13.6\,eV as a function of radius for the AGN, bulge+stellar disk, and ring of NGC\,4138 according to our radiation model. The black line is the total $u$ profile. Solid lines represent the attenuation-free version, and dashed lines represent the attenuated version. Right panel: the fractional contributions of each galaxy component to the total ISRF at a given radius.} 
    \label{fig:ucurve}
\end{figure*}

For both cases, we compute $u_{\mathrm{comp}}$ for each galaxy component at different radii. Each resulting $u_{\mathrm{comp}}$ and the $u_{\mathrm{total}}$ is shown in the left panel of Figure~\ref{fig:ucurve}. When synthesizing D21 PAH spectra, these curves determine the intensity of the radiation to which PAHs are exposed. The right panel of Figure~\ref{fig:ucurve} shows the fractional contribution of each galaxy component to the local energy density at each radius, which informs the relative contributions for the model PAH spectra exposed to different ISRFs. Since the bulge and stellar disk are assigned the same ISRF spectrum, we consider them as a single component for the remainder of our discussion.

\begin{figure}
    \centering
    \includegraphics[width=0.5\textwidth]{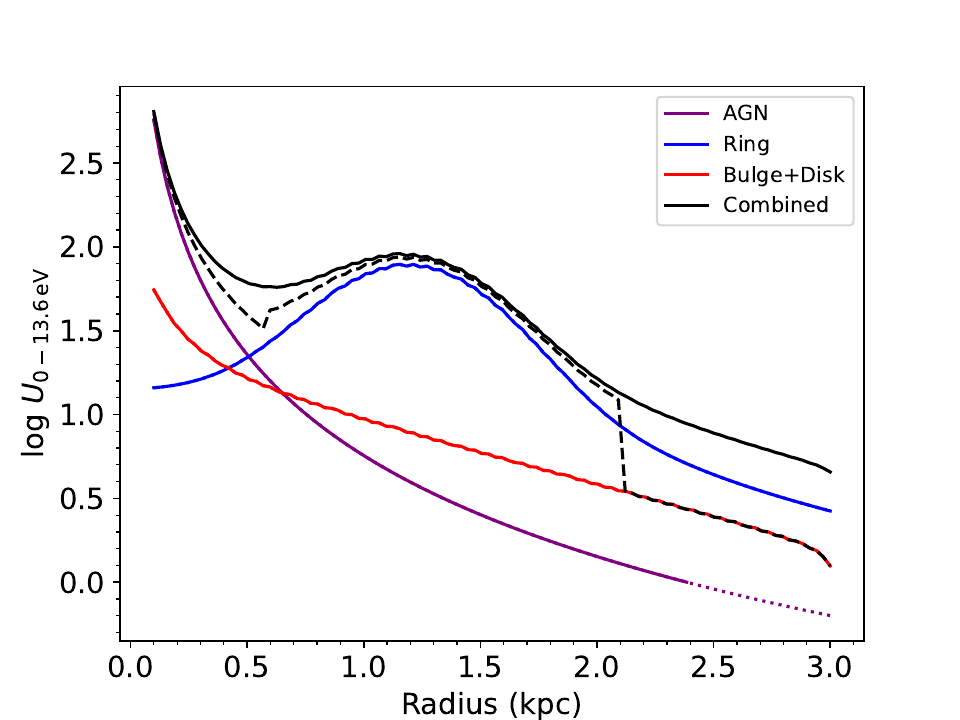}
    \caption{Same as the left panel of Figure~\ref{fig:ucurve}, but for the PAH heating parameter $U$. Dotted segments indicate values of $U$ that are not available in the D21 models. The dashed black line is the combined curve for resulting from the attenuated version. We do not show the component wise curves for the attenuated radiation model for clarity.} 
    \label{fig:capUcurve}
\end{figure}

Finally, the D21 model spectra for different radiation intensities are computed in terms of $U$, so we must convert from $u_{\mathrm{comp}}$ in order to synthesize model PAH spectra. The conversion can be found for each ISRF in D21, see their Eq. 5 and their Table 1. Our model determines a maximum value at the nucleus of $U\approx10^{2.75}$. See Figure~\ref{fig:capUcurve}. 

\section{Results} \label{sec:results}

\begin{figure}
    \vspace*{-0.5 in}
    \includegraphics[width=0.45\textwidth]{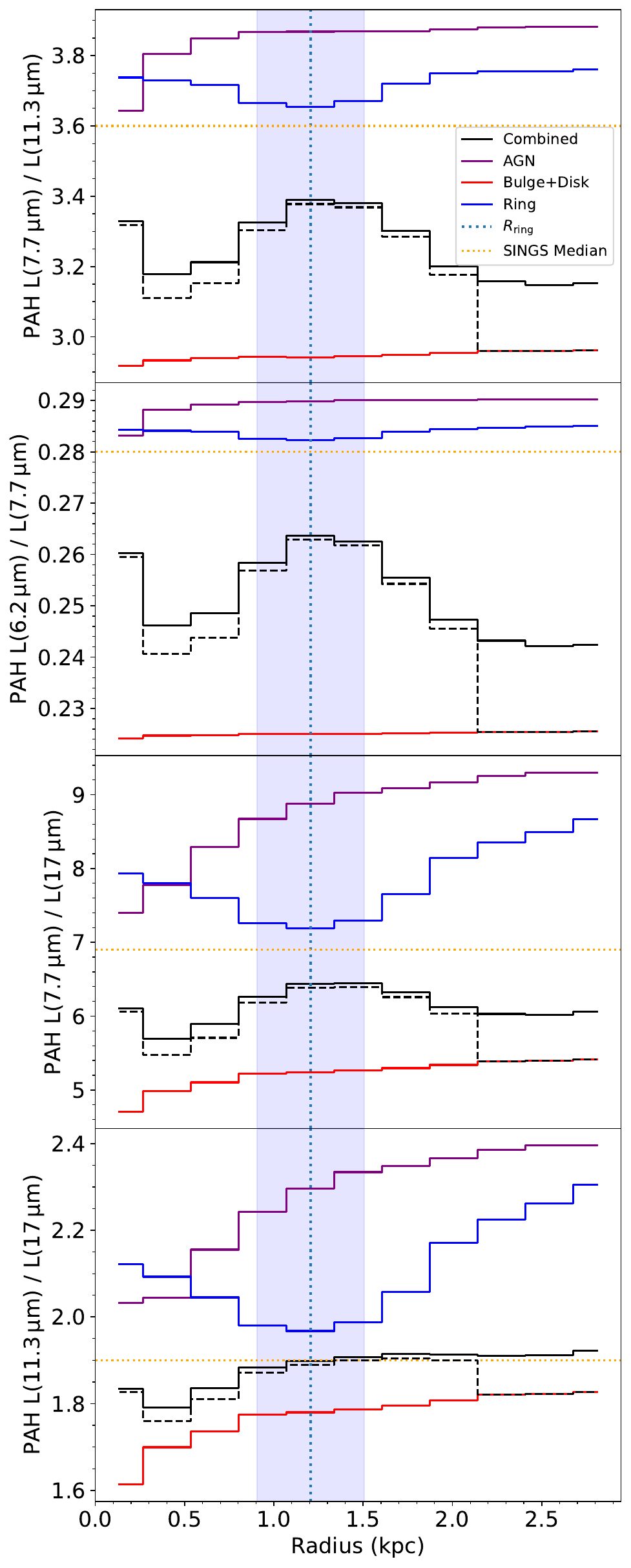}
    \caption{Band ratio profiles predicted by the radiation model when grain size and ionization distributions are held constant, and only the ISRF and $U$ values are changed. Colorful lines represent the profiles predicted from each attenuation-free galaxy component in isolation. The black curves are the weighted-average radial profiles the model predicts using the fractional contributions at each radius from Figure~\ref{fig:ucurve} as weights, where the solid curve is for the attenuation-free model and the dashed curve is attenuated. The fitted $R_{\mathrm{ring}}$ ($1.2\,\mathrm{kpc}$) appears as a vertical light blue dotted line, with the blue shaded area representing $\pm\ 1\,\sigma_{\mathrm{ring}}$.}
    \label{fig:isrf_prediction}
\end{figure}

The D21 PAH models allow us to test the influence of four major parameters on PAH emission. Two of these are extrinsic to PAHs; the average photon energy of light that shines on them ($\left<h \nu\right>$), and the intensity of that light ($U$). Two of these parameters are intrinsic to PAHs; the size of PAH grains, and whether they are ionized or neutral. Together, these extrinsic and intrinsic parameters lend themselves to two testable scenarios. Do AGN-hosts show suppressed short-to-long PAH band ratios because the AGN has an influence on the ISRF of a galaxy? Or, are these band ratios better explained by the photodestruction of smaller PAHs by the AGN and changes to ionization? We attempt to reproduce the observed PAH band ratios at different radii in NGC\,4138 through two different modeling approaches. First, we predict these ratios for a fixed PAH GSD and ionization fraction, varying only the ISRF ($\left<h \nu\right>$, $U$) according to our radiation model. Separately, we hold the ISRF constant and examine PAH band ratios for different GSDs and ionization fractions.

\subsection{Modeling PAH Band Ratios in Varying Radiation Environments} \label{sec:band ratios}

The radiation model developed for NGC\,4138 predicts, at each radius, the integrated 0--13.6\,eV energy density contributed by the spectrum of each galaxy component, shown in Figure~\ref{fig:ucurve}. Since the model gives a maximum $U\approx10^{2.75}$, we assume PAHs everywhere within NGC\,4138 are under the single-photon heating regime (see D21), and therefore we treat each $u_{\mathrm{comp,0-13.6\,eV}}$ as separable; see \S~\ref{sec:discussion} for a discussion of multi-photon heating in NGC\,4138. Thus, the radiation model allows us to test the relative impact of different galaxy components at each radius, for an assumed fixed PAH population. By combining each $u_{\mathrm{comp,0-13.6\,eV}}$ into $u_{\mathrm{total,0-13.6\,eV}}$, this model predicts the observed band ratios at each radius. Both the component-wise and combined radial profiles for four modeled short-to-long PAH band ratios are shown in Figure~\ref{fig:isrf_prediction}. In this ISRF-only scenario, we fix the GSD ($a_{\mathrm{01}}=4\,\mathrm{\AA}$) and ionization fraction ($\mu=10\,\mathrm{\AA}$) at all radii.

For each band ratio profile in the attenuation-free case, the ring spectrum and the AGN spectrum (which we assume is similar in hardness to a 3\,Myr stellar population, see \S~\ref{sec:radiation model}) produce band ratios similar to the S07 SINGS median. The bulge+stellar disk spectrum produces significantly lower short-to-long ratios than those produced by the AGN and ring spectra. This is as expected, as the greater $\left<h \nu\right>$ of the AGN and ring spectra will heat PAHs to higher temperatures with an individual photon than with the lower $\left<h \nu\right>$ of the bulge+stellar disk, and a hotter PAH will tend to emit a greater fraction of energy at shorter wavelengths than at longer wavelengths. Consequently, the bulge+stellar disk is the driver for any short-to-long band ratio suppression when PAH size and ionization distributions are fixed. 
 
The low $\left<h \nu\right>$ of the bulge+stellar disk does produce relatively low short-to-long band ratios at small radii. For most band ratios, these low ratios are not low enough to account for the suppression observed in AGN-hosts. In particular, the modeled $\mathrm{L(7.7\,\mu m)/L(11.3\,\mu m)}$ profile reaches a minimum value around 3.2 at $\mathrm{\sim0.4\,kpc}$, while many AGN-hosts exhibit a $\mathrm{L(7.7\mu m)/L(11.3\mu m)}<2.5$ within the central few kpc, and sometimes $<1$ at the center \citep{smith_07, odowd_09, diamond-stanic_10, garciabernete_22}. 

The shapes of our predicted combined band ratio profiles react strongly to the presence of each galaxy component, according to the changing fractional contribution to the total $u$. Within $r<\mathrm{250\,pc}$, the AGN makes a significant contribution, therefore the band ratios of the predicted combined curves are partway between the AGN and bulge+stellar disk component-wise curves. Thus, our radiation model suggests that the AGN may partially power PAH emission in the nucleus. At the next radius outward, the bulge+stellar disk dominates the fractional contribution to $u_{\mathrm{total,0-13.6\,eV}}$, and band ratios are significantly suppressed here. As the radius increases, the combined band ratio profiles react to the presence of the ring, increasing in value as the fractional contribution of the ring to $u_{\mathrm{total,0-13.6\,eV}}$ increases. At $r>\mathrm{2\,kpc}$, the bulge+stellar disk once again dominates, and band ratios fall again for most profiles.

The profiles that result from the high-extinction radiation model differ somewhat in shape from their attenuation-free counterparts in addition to slightly smaller short-to-long band ratios overall. The radiation originating from the ring does not contribute to the ISRF at small radii ($r<0.57\,\mathrm{kpc}$) for these profiles which results in moderately lower ratios here. Within the ring, band ratios are slightly lower than the unattenuated ratios as the high $\left<h \nu\right>$ radiation produced by the AGN does not help boost them. At radii greater than the extent of the ring ($r<2.1\,\mathrm{kpc}$), band ratios fall substantially as the bulge+stellar disk is the only galaxy component that contributes to the ISRF here.

\begin{figure*}
   \centering
    \includegraphics[width=1\textwidth]{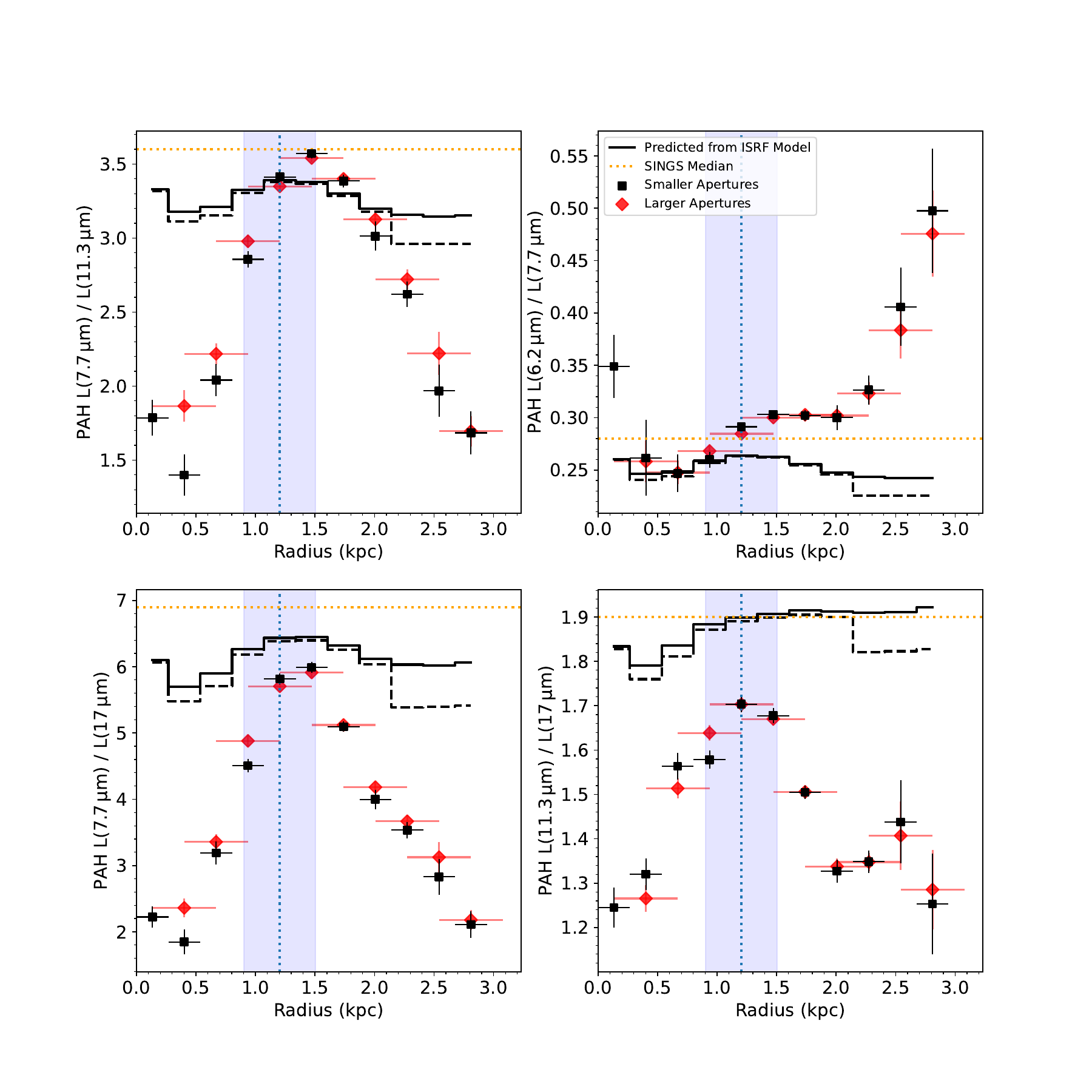}
    \caption{The observed PAH band ratio profiles for the smaller and larger apertures are shown as black squares and red diamonds, respectively. The horizontal bars on the data indicate the span of radii in each aperture. Model-predicted ratio profiles (as in Figure~\ref{fig:isrf_prediction}) are shown as black lines, the attenuation-free model is solid and high-extinction is dashed. The dotted blue line and blue shaded region represent the fitted location of the ring as in Figure~\ref{fig:isrf_prediction}. Median band ratios for the SINGS galaxies in S07 are shown as horizontal orange dotted lines.}
    \label{fig:radial_profiles}
\end{figure*}

\subsection{Observed Band Ratio Profiles} \label{sec:observed ratios}

In Figure~\ref{fig:radial_profiles}, we present the observed short-to-long PAH band ratio profiles in NGC\,4138 alongside the profiles predicted by both versions of the radiation-only model. In shape, the modeled profiles are similar to the observed profiles; they generally have the lowest values where the bulge+stellar disk dominates at $\mathrm{\sim0.4\,kpc}$ and smoothly climb to higher values as they approach $R_{\mathrm{ring}}$, where the ratios are most similar to the SINGS median in S07. As we model the radiation from the ring as being due to a younger stars, this is consistent with \cite{odowd_09}, who found that short-to-long ratios increase near younger stellar populations. In addition, most of the observed band ratio profiles also exhibit an upturn in the central region compared to the adjacent region at $\sim0.4\mathrm{\,kpc}$. Such nuclear upturns are predicted by the combined modeled profiles. The upturn in the $\mathrm{L(6.2\,\mu m)/L(7.7\,\mu m)}$ is still apparent in the ratios from the larger apertures. For the $\mathrm{L(6.2\,\mu m)/L(7.7\,\mu m)}$ and $\mathrm{L(7.7\,\mu m)/L(11.3\,\mu m)}$ profiles, these upturns may be evident as a change in the inflection as the profile approaches the nucleus. Combined, these provide evidence that these rises in short-to-long band ratios are not an artifact of small apertures.

The modeled $\mathrm{L(6.2\,\mu m)/L(7.7\,\mu m)}$, a ratio which is often used as a size indicator for smaller PAHs, exhibits bulge-induced ratio suppression as low as the minimum of the observed profile. For the other band ratios, the observed and modeled profiles differ in two important ways. First, the amplitude of the observed profiles is much greater than the model predicts; while they peak at ratios similar to the SINGS median, they are suppressed far lower than their modeled counterparts. For example, the lowest observed $\mathrm{L(7.7\,\mu m)/L(11.3\,\mu m)}$ is $\mathrm{<1.5}$, which would be among the lowest of the AGN-hosts in the S07 sample. Compare this to a minimum of $\mathrm{\sim3.2}$ for the predicted profiles for the same band ratio. Second, the agreement in the shapes of the modeled and observed profiles is mixed at radii larger than where the ring has its greatest influence. The shape of the attenuation-free modeled $\mathrm{L(7.7\,\mu m)/L(11.3\,\mu m)}$ profile matches the observed quite well at all radii. On the other hand, the $\mathrm{L(7.7\,\mu m)/L(17\,\mu m)}$ and $\mathrm{L(11.3\,\mu m)/L(17\,\mu m)}$ attenuation-free modeled profiles fail to return to the low ratios seen at small radii, but the high-extinction version improves these shapes significantly. Finally, the modeled $\mathrm{L(6.2\,\mu m)/L(7.7\,\mu m)}$ profiles both decrease past this radius instead of the observed increase.


Our bulge-disk-ring decomposition of NGC\,4138 yields $R_{\mathrm{ring}}=1.2\mathrm{\,kpc}$, the radius at which the ring is the brightest. In addition, we have shown that the observed PAH band ratios in NGC\,4138 appear to react to the presence of the ring; in particular, the ring appears to produce higher short-to-long band ratios. Thus, one might expect that the peak of each observed band ratio profile exists at $R_{\mathrm{ring}}=1.2\mathrm{\,kpc}$. However, the peak (local peak in the case of $\mathrm{L(6.2\,\mu m)/L(7.7\,\mu m)}$) in each observed profile is shifted outward from the centroid of the ring, closer to $\mathrm{1.5\,kpc}$.  Intriguingly, the radius at which the ring spectrum makes the greatest fractional contribution to $u_{\mathrm{total,0-13.6\,eV}}$ is also shifted outward according to our radiation model (right panel of Figure~\ref{fig:ucurve}), hence the modeled profiles also peak at radii $>R_{\mathrm{ring}}$. This outward shift is an effect of the bulge mixing with the ring on its inner side.

The modeled ratios involving the $\mathrm{17\micron}$ band do not agree with the observed ratios at any radius: the $\mathrm{17\micron}$ band in NGC\,4138 is enhanced relative to the models, and relative to the SINGS sample. Figure~\ref{fig:radial_profiles} shows that at all radii, the observed band ratios with the $\mathrm{17\micron}$ band in the denominator are well below the SINGS medians. Coupled together, these indicate that the spectra from NGC\,4138 have abnormally strong $\mathrm{17\micron}$ features. This is in agreement with the findings from S07 and \cite{kaneda_08} that AGN-hosts tend to have stronger $\mathrm{17\micron}$ features.

\subsection{Grain Size-Ionization Model Grids} \label{sec:grids}

\begin{figure*}
    \hspace*{-1.7cm}
    \includegraphics[width=1.2\textwidth]{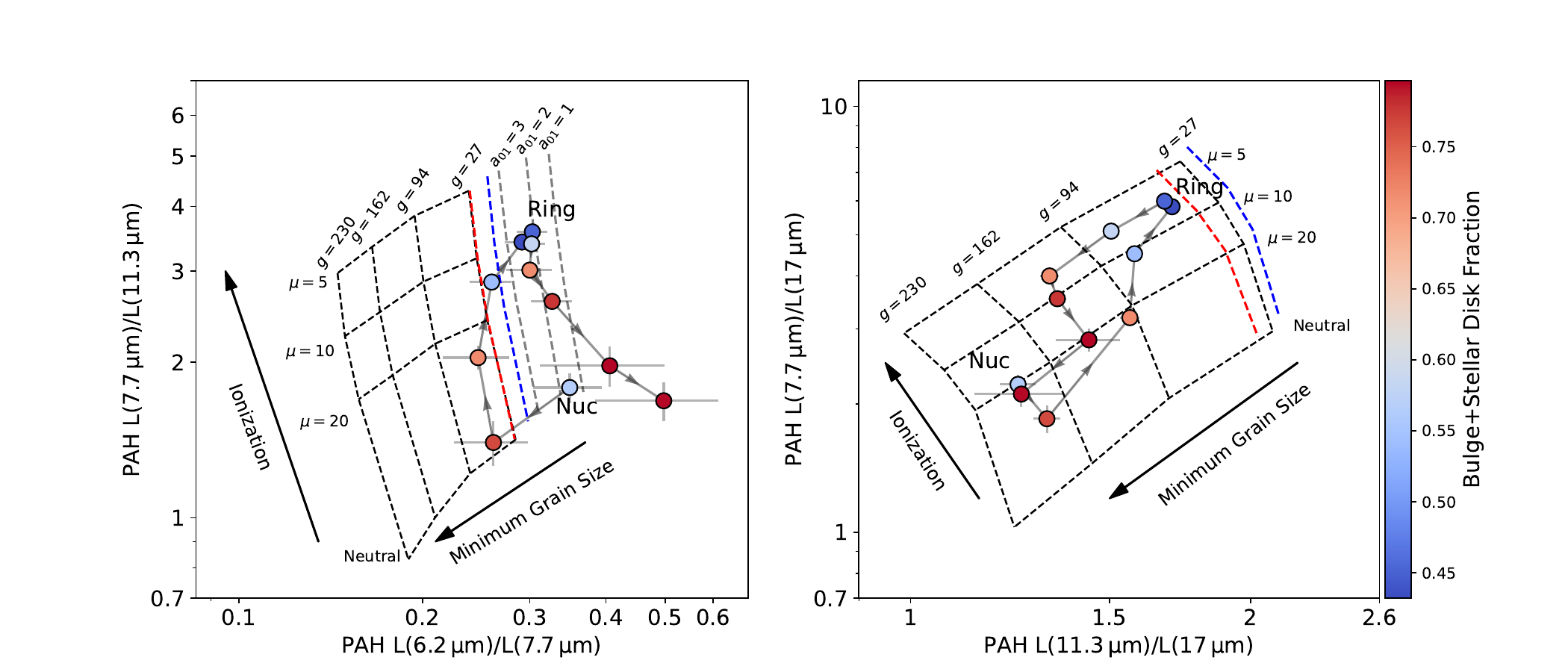}
    \caption{PAH band ratio model grids and the observed radial profiles. Since each galaxy region has a unique ISRF mix, each region produces a unique model grid. Here, we show the median grid from the attenuation-free radiation model in each panel as black dashed lines. For reference, we also show the $g=27$ line for the region which is most bulge-dominated (red dashed line) and the region which is most ring-dominated (blue dashed line). Colored dots are band ratios observed at each radius in the smaller apertures (step size of 270\,pc), with the gray arrows indicating the direction of increasing radius. The colors of the dots indicate the fractional bulge+stellar disk contribution to the ISRF (attenuation free). Values of $\mu$ are in $\mathrm{\AA}$. To the right of the grid in the left panel, we include the $g=27$ line for GSDs with smaller values of $a_{\mathrm{01}}$ (\AA), since the observed track falls off of this side.}
    \label{fig:mu_vs_g}
\end{figure*}

In the previous subsection, we fixed the PAH GSD and ionization fraction, and considered only the variations in PAH band ratios that come with a changing ISRF. We find that the ISRF-only model approximately reproduces the \emph{shapes} of the observed band ratio profiles, but that varying ISRF alone is unable to produce the amplitude of the observed profiles, most of which are suppressed well below the range the ISRF-only model predicts. On the contrary, changes to PAH size and ionization are capable of producing an extremely wide range of possible band ratios. Thus, the suppressed band ratios in AGN-hosts are typically interpreted as being due to alterations of grain properties by an AGN, with the preferential destruction of smaller PAH grains commonly suggested. Here, we explore whether this, for a fixed ISRF, is a sufficient explanation for the shapes of the observed band ratio profiles in NGC\,4138.

We use the D21 model PAH spectra to produce model grids in a PAH band ratio-ratio space, since certain band ratios are more sensitive to PAH ionization ($\mathrm{L(7.7\,\mu m)/L(11.3\,\mu m)}$ and $\mathrm{L(7.7\,\mu m)/L(17\,\mu m)}$), and some are more sensitive to PAH size ($\mathrm{L(6.2\,\mu m)/L(7.7\,\mu m)}$ and $\mathrm{L(11.3\,\mu m)/L(17\,\mu m)}$).  Similar grids are used by D21, and \cite{garciabernete_22} made comparable grids using theoretical PAH spectra computed using Density Functional Theory \citep{rigopoulu_2021}. Our grids are computed for a fixed ISRF, which we choose to be the median ISRF across all radii according to the radiation model. Each vertex in a grid is a unique combination of PAH GSD cutoff ($g$) and average ionization fraction (which is determined by $\mu$), for a GSD with $a_{\mathrm{01}}=4\,\mathrm{\AA}$. This is intended to simulate the destruction of small grains by an AGN (see \S~\ref{background}). Finally, we show the observed band ratio profiles, referred to as ``tracks", plotted on top of the grids. See Figure~\ref{fig:mu_vs_g}.

If the AGN of NGC\,4138 were to be responsible for the variations observed in the band ratio profiles via small grain destruction, then it would have to do so over the multiple kpc scales that we observe these variations over. Further, if changes in $g$ accurately model the effect of an AGN on the GSD, then $g$ should decrease consistently (i.e. smaller PAHs can survive) as the distance from an AGN increases. In the track in the left panel of Figure~\ref{fig:mu_vs_g}, which includes band ratios that are better probes of smaller PAHs, this behavior of $g$ is seen throughout most of the galaxy. The important exception is at the innermost region containing the nucleus ($r<270\,\mathrm{pc}$), where $g$ is expected to be largest, but is nearly at a minimum in this simple interpretation. Since the track in the small grain tracing ratios falls off of the model grid, we include the $g=27$ line for other values of $a_{\mathrm{01}}$, but we emphasize that these produce GSDs with much smaller average PAH sizes than D21 considers. This behavior of $g$ is not shown by the track in the right panel, which primarily probes larger PAHs. The band ratios of these larger PAHs appear to be more influenced by the proximity to the ring for a region rather than proximity to the AGN.

The motion of the observed tracks in both panels of Figure~\ref{fig:mu_vs_g} are similar in the $\mu$ direction of the model grids. Near the nucleus and in the most extended regions, PAHs are more neutral. PAHs are more ionized near the ring, in both cases falling near $\mu=10\,\mathrm{\AA}$, which is the ``standard" ionization distribution in D21. Hence, the ionization fraction of PAHs appears to be a function of proximity to the ring, and the AGN does not appear to influence PAH ionization. Recent \textit{JWST} observations of the AGN-host NGC\,7469 also show no clear link between the AGN and PAH ionization \citep{lai_2022}.

\section{Discussion} \label{sec:discussion}

NGC\,4138 shows significant variations in the observed PAH band ratios at different radii, displaying the diversity of PAH grains and the radiation fields they experience. Overall, we find suppressed short-to-long PAH band ratios near the nucleus and in the bulge for \textit{all} considered ratios, regardless of whether they are more sensitive to PAH size or ionization. Thus, the apparent temperature distribution of PAHs must be changing significantly across NGC\,4138. The two scenarios we consider, changes to radiation or to PAH grain properties, each have merit for explaining the observed variations in PAH emission. However, neither of the two scenarios can comprehensively explain the observed variations themselves, and a complete picture may require considering the effects of both scenarios simultaneously.

The similarity in shape between the observed band ratio profiles and those we predicted with the radiation model suggests that the changing hardness of the radiation field exciting PAH grains ($\left<h \nu\right>$) plays a significant role for the variations in each ratio. We also find that the changing $\left<h \nu\right>$ can explain the radius where the ratio profiles peak, which falls outside of the star-forming ring peak brightness due to the differential effects of bulge heating across the ring, and is a feature not readily explained by the grain destruction hypothesis. However, this is insufficient to produce the amplitude of band ratio variations using the D21 model spectra. The sensitivity of PAHs to changing $\left<h \nu\right>$ in the D21 models are determined by the assumed UV-optical absorption cross-sections of PAHs. Thus, a greater amplitude in our modeled profiles could be attained if PAHs were assumed to be more selective absorbers of harder radiation. 

A greater range of band ratio variation can also be achieved by changing the distributions of intrinsic PAH properties, namely the size and ionization. We see from Figure~\ref{fig:mu_vs_g} that changes to the ionization distribution may supplement radiation effects to achieve the observed amplitudes of the ionization-sensitive $\mathrm{L(7.7\,\mu m)/L(11.3\,\mu m)}$ and $\mathrm{L(7.7\,\mu m)/L(17\,\mu m)}$ profiles; a larger $\left<h \nu\right>$ experienced in the ring may be associated with a greater degree of ionization in PAHs. 

The destruction of small PAHs by the AGN in NGC\,4138 may be the cause of the trend of decreasing $g$ (the lower limit of the PAH GSD) with radius for the track in the left panel of Figure~\ref{fig:mu_vs_g}; this is a commonly suggested mechanism for suppressed short-to-long band ratios observed at the centers of AGN-hosts (see \S~\ref{background}). The clustering of ring-dominated points in the $g$ direction, offset towards smaller $g$, could be attributed to the increased $\left<h \nu\right>$ there. However, it is unclear why the effect of small grain destruction would appear to be strongest at the largest radii, \emph{farthest} from the AGN. In addition, the grain size cutoff appears to be \emph{smaller} near the nucleus, where it is expected to be largest. 

\cite{jensen_17} and \cite{alonso-herrero_14} observed 11.3\,\micron\ PAH emission within 10's of pc from several LLAGN-hosts, and found that PAH surface brightness was stronger toward the LLAGN. This could be evidence that the abundant UV photons emitted from an AGN may be able to photo-excite PAHs in the vicinity. This could be the mechanism causing the apparently smaller $g$ and the relatively larger short-to-long band ratios observed near the nucleus compared to the adjacent region at $\sim 0.4\,\mathrm{kpc}$. The right panel of Figure~\ref{fig:ucurve} shows that the AGN contributes significantly to the total share of the 0--13.6\,eV energy density at the nucleus, causing the nuclear ``upturns" in the modeled profiles. If the hard radiation starburst spectrum is a good approximation of the filtered AGN light that PAHs are exposed to, then the observed upturns may be consistent with AGN-powered PAH emission.

However, we are limited by the spatial resolution of \textit{Spitzer}/IRS with $\mathrm{PSF\ FWHM}\approx4\,\arcsec$, and our central region has a radius of 270\,pc. Thus, the nucleus is unresolved and this region is heavily contaminated by PAH emission in the central $0.5\,\mathrm{kpc}$, which makes any evidence for AGN-powered PAH emission ambiguous. Indeed, an unresolved star-forming region within this radius would induce similar effect. Additionally, the spectrum of the region near the nucleus is the most challenging for quantifying PAH strengths due to dilution from the AGN continuum, absorption features in starlight, and possible emission from silicate grains.

With an angular resolution far superior to \textit{Spitzer}, \textit{JWST} will provide further insights into this issue. Early \textit{JWST} observations of PAH band ratios in the AGN-host NGC\,7469 ($D=70.6\,\mathrm{Mpc}$) were shown by \cite{lai_2022} to exhibit a similar trend of decreasing $\mathrm{L(6.2\,\mu m)/L(7.7\,\mu m)}$ with decreasing radius (see also: \cite{Asmus2023}). Intriguingly, they too detect a much larger $\mathrm{L(6.2\,\mu m)/L(7.7\,\mu m)}$ near the nucleus analogous to the upturn in our ratio profile. In the same galaxy, \cite{lai_23} detected an enhancement in $\mathrm{L(7.7\,\mu m)/L(11.3\,\mu m)}$ near the nucleus, another short-to-long ratio. \cite{garciabernete_7469} also reports a larger relative $\mathrm{L(6.2\,\mu m)/L(7.7\,\mu m)}$ in the nucleus using the same observations. However, the behavior of $\mathrm{L(7.7\,\mu m)/L(11.3\,\mu m)}$ is seemingly different between the two studies; \cite{lai_2022} shows a increasing trend toward the nucleus, whereas \cite{garciabernete_7469} reports the smallest ratio there. More \textit{JWST} observations of the nuclei of closer AGN-hosts are required, and would provide an even better opportunity to observe possible PAH excitation by the AGN, in addition to probing the impact of the AGN on the PAH GSD.

The larger PAHs ($N_{\mathrm{C}} \gtrsim 500$) tell a different story in Figure~\ref{fig:mu_vs_g}. Instead of a generally decreasing $g$ across all radii, the tracks of $\mathrm{L(7.7\,\mu m)/L(11.3\,\mu m)}$ vs. $\mathrm{L(11.3\,\mu m)/L(17\,\mu m)}$ for larger PAHs indicate a decrease in $g$ only up to $\sim\,R_{\mathrm{ring}}$, before turning around and returning to implied values of $g$ similar to near the nucleus. This behavior can not be explained via AGN grain destruction, since it implies that the AGN exerts a greater influence on the PAH GSD at $>R_{\mathrm{ring}}$ than it does at $\approx R_{\mathrm{ring}}$. Rather, it suggests that the larger PAHs are reacting to conditions in the ring that are different compared to the bulge+stellar disk. These conditions may affect the GSD via grain destruction (e.g., sputtering in hot plasma), or they may involve a boost to the impact of radiation on PAH band ratios. 

One such boost could come from a different treatment of $U$ in our radiation model. \cite{aniano_2020} produced spatially-resolved maps of the radiation heating intensity $U$ for the galaxies in the KINGFISH sample \citep{KINGFISH}, and although they were limited by the resolution of the \textit{Spitzer} MIPS160 and the \textit{Herschel} SPIRE250 cameras ($\mathrm{PSF\ FWHM}=38\farcs8\mathrm{\ and\ }18\farcs2\,\mathrm{,\ respectively}$), they estimated that between 5--20\% of the dust emission of a typical galaxy within a given pixel ($\mathrm{MIPS160}=18\,\arcsec$, $\mathrm{SPIRE250}=6\,\arcsec$) was due to dust near photodissociation regions (PDRs), heated by $U>100$. Above this threshold, PAHs begin to experience multi-photon heating, in which the radiative relaxation time for a PAH is less than the time between incident photons. For given incident photon energy, larger PAHs have both a larger absorption cross section and longer radiative cooling time than smaller PAHs \citep{draine_07}, hence larger PAHs enter the multi-photon heating regime at lower values of U compared to smaller PAHs. Consequently, short-to-long PAH band ratios become sensitive to $U$ for $U>100$ (see, e.g., D21 Figure~19). 

\begin{figure}
   \centering
    \includegraphics[width=0.5\textwidth]{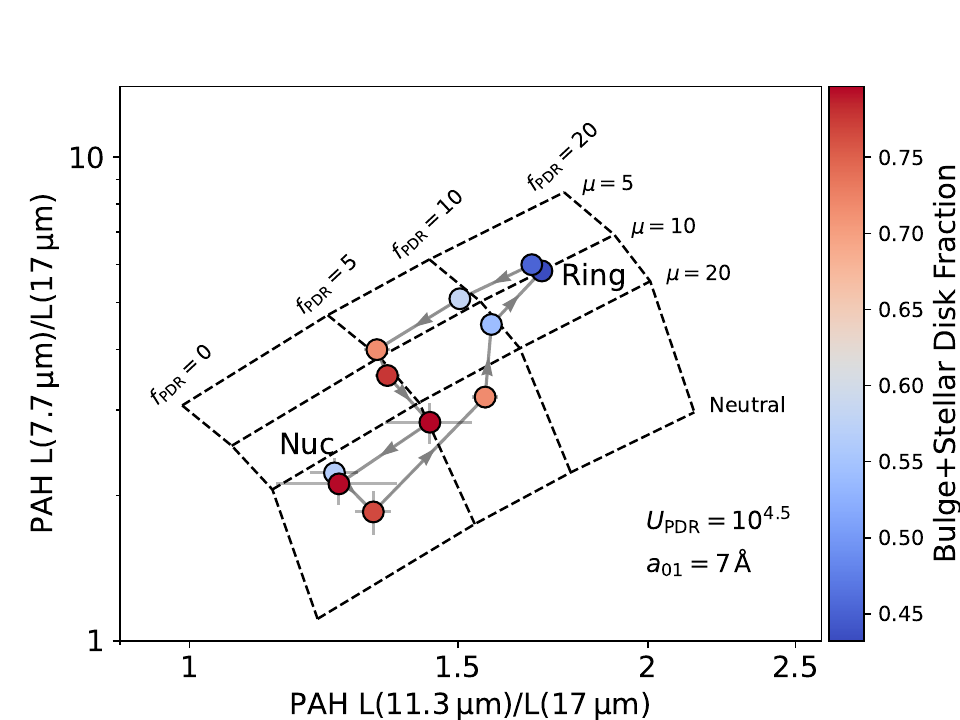}
    \caption{Same as the right panel of Figure~\ref{fig:mu_vs_g}, but the size direction (previously denoted by different $g$ values) has been replaced with varying $f_{\mathrm{PDR}}$ (\%).}
    \label{fig:mu_vs_fPDR}
\end{figure}

In the absence of such a $U$ map for NGC\,4138, we made the assumption that PAHs are heated primarily by radiation in the diffuse ISM, within the single-photon heating regime. A boost to $U$ provided in some regions by exposure of PAHs to nearby PDRs would not only make our radiation model more physically realistic, but it would increase the amplitudes of the modeled ratio profiles in the radiation-only scenario. 

Figure~\ref{fig:mu_vs_fPDR} is similar to the right panel of Figure~\ref{fig:mu_vs_g}, but the $g$ direction of the model grid has been replaced with $f_{\mathrm{PDR}}$, the fraction of PAH luminosity heated by $U_{\mathrm{PDR}}$, which we define as the ring-ISRF heating experienced by PAHs near a PDR. We set $a_{\mathrm{01}}=7\,\mathrm{\AA}$ for the GSD of the model grid so that the $f_{\mathrm{PDR}}=0\,\%$ line has lower short-to-long ratios than the observed track; this accounts for the abnormally large $17\micron$ feature observed in all regions of NGC\,4138, although it describes a GSD with a substantially larger average PAH size than is discussed in D21. $U_{\mathrm{PDR}}$ is chosen to fit the following criteria: 1) it produces a model grid which spans the data given the typical range of $f_{\mathrm{PDR}}$ in \cite{aniano_2020} (typically up to 20\%) and 2) the range of the ionization parameter $\mu$ covered by the observed track is similar to that of the same track in the right panel of Figure~\ref{fig:mu_vs_g}. We find that $U_{\mathrm{PDR}}=10^{4.5}$ satisfies these conditions.

Using this example, Figure~\ref{fig:mu_vs_fPDR} shows that it is possible to produce the observed range of band ratios for larger PAHs by varying only their ionization and starlight heating intensity, rather than alterations to the GSD. In this conceptual framework, the observed track in Figure~\ref{fig:mu_vs_fPDR} suggests that $f_{\mathrm{PDR}}$ is greater in regions with a larger influence from the ring ISRF, and it is lower in bulge and stellar disk-dominated regions. The true $a_{\mathrm{01}}$ and $U_{\mathrm{PDR}}$ in NGC\,4138 are unknown, but this example illustrates that the small amplitudes of the radiation-only modeled band ratio profiles for larger PAHs might be due in part to an unrealistic treatment of $U$.

To place the model grid in Figure~\ref{fig:mu_vs_fPDR} on the observed track requires a GSD with $a_{\mathrm{01}}=7\,\mathrm{\AA}$. For reference, the ``large" GSD discussed in D21 has $a_{\mathrm{01}}=5\,\mathrm{\AA}$. This is not necessarily of concern on its own, since our decision to consider the canonical $a_{\mathrm{01}}=4\,\mathrm{\AA}$ for NGC\,4138 was not made from any observational consideration. However, such an ``extra large" GSD would exacerbate the issue for the smaller PAHs in Figure~\ref{fig:mu_vs_g}, which calls for a much smaller $a_{\mathrm{01}}$ to contain the observed track within the model grid; equivalently, the modeled $\mathrm{L(6.2\,\mu m)/L(7.7\,\mu m)}$ profile would become far too suppressed compared to the observed profile. Smaller PAHs are affected by multi-photon heating, but not by enough to significantly impact ratios at $U_{\mathrm{PDR}}=10^{4.5}$. A $U$ map of NGC\,4138, combined with the modeled variations of $\left<h \nu\right>$, might help to resolve this discrepancy.

\section{Summary} \label{sec:summary}

As PAHs become increasingly used as indicators of star formation and the conditions in the ISM of galaxies, there is an urgency to understanding the mechanism of PAH suppression in AGN hosts. Though an interaction between the AGN and PAHs is often suggested as the cause, an alternative explanation is the rapidly varying radiation environment away from the AGN center, stellar bulge, and star-forming disk. Disentangling these potentially competing effects has proven difficult. In this study, we attempt to separate the effects of the starlight and those of the AGN by modeling the ISRF of star-forming ring galaxy NGC\,4138.  Drawing from the new PAH emission models of \cite{draine_21}, the radiation model is used to predict the influence of distinct ISRF components on PAH band ratios across the galaxy. We compare these predicted band ratios with those of observed spectra at many radii using a large \textit{Spitzer}/IRS spectral map. The main conclusions from this work are the following:

\begin{itemize}

    \item The observed short-to-long PAH band ratios in NGC\,4138 are variable with distance from the nucleus. These ratios are suppressed at small radii, and smoothly increase to values typical for star-forming galaxies near the star-forming ring. Outside of the ring, the ratios typically fall back to values seen at small radii, indicating that the AGN does not dictate the band ratios at $\gtrsim$kpc scales.
    
    \item Our ISRF model suggests that varying radiation environments contribute to the changing band ratios, yielding similar trends across the outer disk, star-forming ring, and central bulge. Incorporating attenuation effects further improves the agreement in these trends. However, the model under-predicts the amplitude of these variations by up to $\sim2.5\times$.
    
    \item Significant changes in PAH UV-Optical opacity to increase the sensitivity of PAH grains to UV photons would be needed to reproduce the large amplitude of band ratio variations.  Increased PAH heating from higher radiation intensity near PDR surfaces in the star-forming ring could boost short-to-long band ratios for larger PAHs, but this can reproduce the observed ratios only by assuming a substantially larger size distribution than is typically considered.

    \item Modification of the PAH population by the AGN via small grain destruction and bulk changes in ionization state could account for some of the observed variation in band ratios, although models incorporating these changes lack consistency between short and long PAH bands. Based on central upturns in band ratios, the AGN may also directly influence PAH emission within a few hundred pc of the nucleus.
    
\end{itemize}

An extended analysis on a much larger sample will facilitate a more complete picture of the relationship between PAH band ratios and the radiation environments in galaxies, and the possible direct relationship between PAH emission and AGN. As it had the ability to efficiently map entire galaxies, \textit{Spitzer} observations will continue to be essential for the understanding of PAHs at large spatial scales, but \textit{JWST} is the first facility capable of observing the full suite of PAH features in AGN host galaxies at scales of 10's of pc. These observations can include the grain size-sensitive $3.3\,\micron$ PAH feature, which will likely be critical for understanding variations in grain size distributions at the smallest sizes, as well as radiation effects on PAHs. Future \textit{JWST} observations of the nuclei of AGN hosts may provide the first direct confirmation of the role AGN play on the physical characteristics of PAH grains.

\section{Acknowledgements} \label{sec:acknowledgements}

We thank the members of R. Chandar's Paper Writing Class (2022-2023) for the constructive feedback over many iterations. We also thank B. Hensley and the attendees of the 2023 PAH Fest for helpful suggestions and conversations. We thank the anonymous referee for their insightful comments which improved this work. We gratefully acknowledge support for this project from the Research Corporation for Science Advancement through Cottrell SEED Award No. 27852. This research was supported in part by NSF grant AST-1908123. This research has made use of the NASA/IPAC Infrared Science Archive, which is funded by the National Aeronautics and Space Administration and operated by the California Institute of Technology.  JDS gratefully acknowledges support for this project from the Research Corporation for Science Advancement through Cottrell SEED Award No.
27852.  This research has made use of the NASA/IPAC Extragalactic Database (NED), which is funded by the National Aeronautics and Space Administration and operated by the California Institute of Technology.

\facilities{\textit{Spitzer}/IRS (AORKEY: 18240768), \textit{Spitzer}/IRAC and MIPS Super Mosaics \citep{supermosaics}, \textit{GALEX} (DOI: 10.17909/T9H59D) \citep{galex_data}}

\software{Astropy \citep{astropy:2013, astropy:2018, astropy:2022}, Matplotlib \citep{Hunter:2007}, NumPy \citep{harris2020array}, SciPy \citep{2020SciPy-NMeth}}

\bibliography{citations}{}
\bibliographystyle{aasjournal}

\end{document}